\def\epsfpreprint{Y}   

\def\PRD{N}            

\if \epsfpreprint Y
\documentstyle[12pt,epsf]{article} \fi
\if \epsfpreprint N
\documentstyle[12pt]{article} \fi

\def\figsizeA{6.2}
\def\figsizeB{6.0}

\def\figsizeD{5.6}

\def\figure#1#2#3{\if \epsfpreprint Y \epsfxsize=#3truein
\centerline{\epsffile{fig_#1.eps}}
\centerline{\vbox{{\bf \noindent Figure #1.} #2}}
\bigskip \fi}

\def\Psibar{\overline{\Psi}}
\def\slash{\!\!\!/}
\def\bfm#1{\mbox{\boldmath $#1$}}
\def\MeV{{\rm\  MeV}}
\def\GeV{{\rm\  GeV}}
\def\b{\tilde{\beta}}
\def\kpnu{k_\nu \!+\! {q_\nu \over 2}}
\def\kmnu{k_\nu \!-\! {q_\nu \over 2}}
\def\kp{k \!+\! {q \over 2}}
\def\km{k \!-\! {q \over 2}}
\def\fs{\ \ \ .}

\def\spose#1{\hbox to 0pt{#1\hss}}
\def\ltapprox{\mathrel{\spose{\lower 3pt\hbox{$\mathchar"218$}}
 \raise 2.0pt\hbox{$\mathchar"13C$}}}
\def\gtapprox{\mathrel{\spose{\lower 3pt\hbox{$\mathchar"218$}}
 \raise 2.0pt\hbox{$\mathchar"13E$}}}
\def\inapprox{\mathrel{\spose{\lower 3pt\hbox{$\mathchar"218$}}
 \raise 2.0pt\hbox{$\mathchar"232$}}}

\def\one{$r=0$, $N_c=3$, $F_\pi=93 \MeV$, and $M_\pi=140 \MeV$. From
top to bottom the lines correspond to the $\sigma$ mass, quark mass
and $\sigma$ width calculated in the large $N$ approximation to
leading order in $m_q$. The vertical line denotes the point where the
quark mass is equal to $310 \MeV$.}
\def\twoA{The $\beta_1$, $\kappa = 1 / (8 +2 m_0)$ plane for
$r=1.0$, $N_c=2$. The solid lines are constant $m_\pi$ lines.  From
top to bottom they correspond to $m_\pi=0, 0.1, 0.2, 0.3$. The dotted
lines are constant $m_q$ lines. From right to left they correspond to
$m_q=-0.2,-0.1,0.0$, $0.1$, $0.2$, $0.3$. The $m_q=0, m_\pi=0$ point
is located at $\beta_{1_{chiral}} =0.3416$, $\kappa_{chiral}=0.3994$.}
\def\twoB{Same as in Figure 2a but for $r=0.1$. The $m_q=0$,
$m_\pi=0$ point is located at $\beta_1=2.1492$, $\kappa=34.843$.}
\def\three{$r=0$, $m_0=0$, $N=2$, $L_x=8$, $L_t=16$.  The diamonds are
the values of $m_q$ and the crosses are the values of
$m^{\prime}_q=\{-N <\bar\Psi\Psi> / (2 \beta_1)\} + m_0$ from the
numerical simulation. Because of the functional identity, eq.
\ref{fun_identity}, the two quantities are expected to be equal.  The
solid line is the large $N$ prediction on the same size lattice. The
dotted line denotes the infinite volume $\beta_{1_c}$ from the large
$N$ calculation. For $\beta_1\le\beta_{1_c}$ the model is in the
broken phase.}
\def\four{Same as in Figure 3 but for the $U(2) \times U(2)$ NJL and
for $r=1$, $L_t=8$. Notice that the diamonds and crosses are almost
identical.}
\def\five{$r=1$, $\beta_1=2.5$, $N=2$. The diamonds are the values
of $m_q$ from the numerical simulation for $L_x=8$, $L_t=16$. The solid
line is the large $N$ result on the same size lattice.}
\def\six{{\bf a)} The $\sigma$ propagator in momentum space for $10$
small momenta with $r=0$, $m_0=0$, $\beta_{1_{MC}}=2.2$, $\langle
\sigma \rangle = \sigma_s = 0.4840$, $N=2$, $L_x=16$, $L_t=16$. The
crosses are the values from the numerical simulation.  The diamonds
are the large $N$ results and ``fit'' the numerical results with a
$\chi^2 / d.o.f. = 0.38$. {\bf b)} Same as in 6a but for
$\beta_{1_{MC}}=2.4$, $\langle \sigma \rangle = \sigma_s = 0.35$. The
large $N$ results ``fit'' the numerical results with a $\chi^2 /
d.o.f. = 0.67$.  {\bf c)} Same as in Figure 6a but for the pion
propagator. The large $N$ results ``fit'' the numerical results with a
$\chi^2 / d.o.f. = 0.32$. {\bf d)} Same as in 6b but for the pion
propagator. The large $N$ results ``fit'' the numerical results with a
$\chi^2 / d.o.f. = 0.42$.}
\def\sevenA{$\beta_1 = 2.5$, $r=1$, $N=2$, $L_x=8$, $L_t=16$.  The
crosses are the MC data. The solid line is the large $N$ prediction on
the same size lattice.}
\def\sevenB{Same as in Figure 7a but as a function of $\langle \sigma
\rangle$.}
\def\eight{$r=0$, $m_0=0$, $N=2$.  The diamonds are the values of
$m_q$ from the numerical simulation for $lx=8$, $lt=16$.  The solid
lines are the large $N$ numbers with the zero mode included. From
right to left they correspond to $(L_x=8,L_t=16)$, $(L_x=16,L_t=16)$,
$(L_x=32,L_t=32)$, $(L_x=64,L_t=64)$. The dotted lines are the large
$N$ numbers with the zero mode excluded. From right to left they
correspond to $(L_x=16,L_t=16)$, $(L_x=8,L_t=16)$.  The solid vertical
line denotes the infinite volume $\beta_{1_c}$ from the large $N$
calculation. For $\beta_1\le\beta_{1_c}$ the model is in the broken
phase.}
\def\nine{$r=1$, $\beta_1=2.5$, $N=2$, $L_x=8$, $L_t=16$. The lines
are the large $N$ effective potential with $m_0=-0.2$ (bottom) and
$m_0 =-0.5$ (top). The crosses indicate the absolute minimum.
$V_{eff}$ is infinite at the spikes.}
\def\tenA{$r=0$, $m_0=0$, $N=2$, $L_x=16$, $L_t=16$.  The solid
lines are the real part of the inverse $\sigma$ propagator for
external $4$--momentum $q=(i m_\sigma, 0, 0, 0)$ at large $N$ with the
zero modes included. From left to right they correspond to $\beta_1 =
2.3, 2.25, 2.2$ or equivalently to $\sigma_s = 0.278, 0.323, 0.366$.
The functions have discontinuities that are denoted with the doted
lines.  The star denotes the relevant zero for the $\beta_1=2.2$
line.}
\def\tenB{$r=0$, $m_0=0$, $N=2$, $\beta_1=2.3$ The lines are the
real part of the inverse $\sigma$ propagator for external
$4$--momentum $q=(i m_\sigma, 0, 0, 0)$ at large $N$. The dotted line
is for $L_x=16$, $L_t=16$ with the zero mode included, the solid line
for $L_x=32$, $L_t=32$ with the zero mode included, the dot--dash line
is for $L_x=16$, $L_t=16$ with the zero mode excluded, and the dashed
line is the infinite volume result from the asymptotic expansion. The
stars denote the relevant zeros.}
\def\elevenA{$r=1$, $N=2$, $L_x=8$, $L_t=16$. The solid line denotes
the $m_\pi=0$ line.}
\def\elevenB{Detail around the ``prong" that corresponds to $m_q=0$
(``prong" ``3" of figure 11a).}
\begin{document}
\if \PRD Y \baselineskip = 2\baselineskip \fi

\title{\bf A Study of the Nambu--Jona-Lasinio Model on the Lattice}
\vskip 1. truein
\author{Khalil M. Bitar and Pavlos M. Vranas \\
Supercomputer Computations Research Institute \\
The Florida State University \\
Tallahassee, FL 32306}
\maketitle


\begin{abstract}
\if \PRD Y \baselineskip = 2\baselineskip \fi

We present our full analysis of the two flavor Nambu--Jona-Lasinio
model with $SU(2) \times SU(2)$ chiral symmetry on the
four--dimensional hypercubic lattice with naive and Wilson fermions.
We find that this model is an excellent toy field theory to
investigate issues related to lattice QCD. We use the large $N$
approximation to leading order in $1/N$ to obtain non perturbative
analytical results over almost the whole parameter range. By using
numerical simulations we estimate that the size of the $1/N$
corrections for most of the quantities we consider are small and in
this way we strengthen the validity of the leading order large $N$
calculations. We obtain results regarding the approach to the
continuum chiral limit, the effects of the zero momentum fermionic
modes on finite lattices and the scalar and pseudoscalar spectrum.

\end{abstract}

\vfill
PACS 12.40.-y, 11.30.Rd, 12.38.Gc \hfill
\newpage

\section{Introduction}

At present the low energy properties of QCD can only be studied
numerically using lattice gauge theory. It is believed that in order
to obtain reasonable results in reasonable time periods the computing
power needed is of the order of one Teraflop. Although supercomputers
with such capacity may be built in the next few years it is still
important to use other methods to get insights into the physics as well
as into the behavior of the theory on the lattice.

The Nambu--Jona-Lasinio (NJL) model was introduced before the
discovery of quarks as a theory of nucleons interacting with a
four--Fermi interaction. Today the fermionic fields of the model are
reinterpreted as being those of the quarks. The most important feature
of the model that qualifies it to describe some of the important low
energy properties of QCD is that it possesses the same chiral symmetry
as QCD and that this symmetry can be realized in the Goldstone mode.

The NJL model can also be motivated by an argument found in
\cite{Dhar-Shankar-Wadia}. The gauge field in the full theory of QCD
develops a finite correlation length of the size of the inverse mass
of the lightest glueball ($\approx 1550 \MeV$, see for example
\cite{UKQCD}). If we consider QCD on a lattice and integrate the high
momentum fluctuations of the fermionic and gauge fields down to
$\Lambda \ltapprox 1550 \MeV$ then the lattice spacing will be of the
order of the correlation length and we must then essentially have a
theory of fermions with contact interactions and cutoff $\Lambda$. The
resulting effective Lagrangian will maintain the original chiral
symmetry but will of course be more complicated. If we further
restrict our attention to energies much below the cutoff, naively
speaking, it should be enough to keep in the Lagrangian the least
irrelevant operator namely the four--Fermi dimension six operator. This
is the NJL model.

Unfortunately, by only keeping the four--Fermi operator, valuable
information was lost and the model does not confine the quarks.
Therefore, strictly speaking, it can not be a true effective field
theory of QCD.  Furthermore, if, for example, we want to study the
$\sigma$ particle, which on phenomenological grounds is believed to
have mass $\approx 750 \MeV$, then the separation of scales is
probably not large enough to justify the neglect of operators with
dimension higher than six.  Nevertheless, as mentioned above, the NJL
model possesses the same chiral symmetry as QCD and it can realize
this symmetry in the Goldstone mode. It is this feature that is most
crucial in the understanding of the lightest hadrons and the reason
for the successful quantitative predictions of the model.

The NJL model has been studied extensively for various cases with
continuum type regularizations. For a comprehensive review the reader
is referred to \cite{Klevansky} and references therein. Also for some
recent work in the continuum see \cite{recent}. Furthermore the NJL
model is a special case of Yukawa models that, under a different
context, have been studied extensively with lattice regularization
\cite{Yukawa}. The model has also been studied on the lattice
\cite{Hasenfratz} in connection with the possible equivalence of the
top quark condensate with the Higgs field \cite{Bardeen}.  In that
work, however, the separation of scales is very large (the cutoff is
of the order $10^{14} \GeV$), and it is therefore quite a different
problem than the one considered here.

In this paper we do not attempt to use the NJL model to make physical
predictions. Our interest in the NJL model is purely qualitative and
originates from the fact that its lattice version is an excellent toy
model to investigate issues related to lattice QCD. This is the case
primarily because it is an effective four dimensional fermionic theory
that has the same chiral symmetry as QCD, which symmetry in an
appropriate phase may be realized in the Goldstone mode, and because
one can use the large $N$ approximation to leading order in $1/N$ to
obtain non perturbative analytical results over almost the whole
parameter range. Also by using numerical simulations of the model one
can estimate the size of the $1/N$ corrections and in this way
strengthen the validity of the leading order large $N$ calculations.
As it will be shown in this paper the $1/N$ corrections estimated in
this way are small for most of the quantities we consider. This means
that one has a good analytical handle on the model. Because of this a
wealth of information can be extracted.

For naive fermions, we calculate the scalar and pseudoscalar spectrum
as a function of the cutoff and find agreement with general
expectations. This is not perceived as a quantitative prediction but
rather as a first approximation to a physical picture.  Furthermore,
since the NJL model possesses the right symmetry of a possible
embedding theory of the Higgs Sector it can therefore provide an
interesting example of such a theory. We briefly discuss this
possibility.

One of the main purposes of our work involves the study of the Wilson
formulation of lattice fermions. This formulation breaks the chiral
symmetry of the model explicitly. The symmetry is expected to be
restored as a result of the tuning of the Wilson parameter $\kappa$ to
some critical value $\kappa_c$. Needless to say this is a very
important issue for QCD and it would be very useful to clarify how it
is realized in a simpler model. This involves, for example, issues
such as what the continuum chiral limit on a lattice is and how does
one define $\kappa_c$. For Wilson fermions we also determine the
effect of the heavy doublers on the sigma particle. We show that the
heavy doublers raise the mass of the sigma to be of order cutoff.

Another issue investigated in this paper involves the study of the
effects of the zero modes on finite lattices. This is of particular
importance to numerical simulations. We demonstrate that in certain
cases these effects are large and can obscure the extrapolation to
infinite volume physics. We also estimate the size of the $1/N$
corrections by comparing leading order large $N$ results on finite
size lattices with numerical results on the same size lattices and
find that in most cases the $1/N$ corrections are small. This is of
interest since, as mentioned above, the NJL model has been and is
studied extensively as a phenomenological model using continuum type
regularizations and among other methods the leading order large $N$
approximation.

In this paper we consider the two flavor (up and down) NJL model with
$SU(2) \times SU(2)$ chiral symmetry and SU(N) color symmetry, with
scalar and pseudoscalar couplings \cite{Ebert-Volkov} on the four
dimensional hypercubic lattice. We consider both naive and Wilson
fermions and we study the model using a large $N$ expansion as well as
a Hybrid Monte Carlo (HMC) numerical simulation. A shorter version of
this work containing only the main results has appeared elsewhere
\cite{Bitar-Vranas}.

Throughout the paper we use a notation whereby small letters denote
quantities in lattice units and capital letters denote quantities
in physical (MeV or GeV) units. The lattice spacing is denoted by
``$a$".

The paper is organized as follows. The model and its lattice version
are described in Section 2. The large $N$ analysis to leading order in
large $N$ is given in Section 3. Using asymptotic expansions we study
in Section 3.1 the case of naive fermions on an infinite lattice and
present results regarding the scalar--pseudoscalar sector of QCD and
briefly discuss their possible relevance to the Higgs sector. In
Section 3.2 we study the case of Wilson fermions on an infinite
lattice using asymptotic expansions and present results regarding the
approach to the chiral continuum limit as well as the effect of the
doublers on the scalar particle self energy. The numerical and large
$N$ work on finite lattices is described in Section 4. We present
there, in Section 4.1, a comparison of our numerical results with the
large $N$ results obtained on same size lattices and obtain an
estimate of the size of the $1 / N$ corrections. In Section 4.2 we
examine the effects of the zero momentum quark modes on finite
lattices in connection to the inversion time of the Conjugate Gradient
algorithm used in the HMC, and more importantly in connection to
finite size effects.  Furthermore after the preprint version of this
work appeared in the computer list ``hep-lat" and during the
refereeing process of this paper another group \cite{Aoki} presented
complementary results. In particular, for the case of Wilson fermions,
the authors showed, using the leading order large $N$ approximation,
that there is a phase where the remaining flavor-parity symmetry
breaks spontaneously generating a non zero vacuum expectation value
for the pion field. Their work was done at infinite volume using
numerical integration and among other things it was shown that the
phase line is a line at which all three pions are massless. This line
was plotted for all effective quark masses. In this work we always
stay on the symmetric side of this line where the vacuum expectation
value of the pion field is zero. At infinite volume, using asymptotic
expansions, we had calculated the part of this line (fig. 2a) that
corresponds to small effective quark masses since this is the region
where continuum physics is extracted. Our calculations, up to errors
relevant to the approximations used, are in agreement.  However, on a
finite volume the comparison is not as straight forward and it
deserves special attention. For this reason section 4.3 was added
where the zero pion mass line on a finite volume is discussed in
detail. Finally, a short summary and conclusion is given in Section 5.

\section{The model}

The Lagrangian density in Minkowski space and in continuum notation
is:
\begin{equation}
{\cal L} = \Psibar(i \partial\slash  - m_0)\Psi +
{G_1 \over 2}\left[(\Psibar\Psi)^2 +
(\Psibar i \gamma_5 \bfm{\tau} \Psi)^2 \right].
\label{Langrangian1}
\end{equation}
In the above expression all indices have been suppressed. The fermionic
field $\Psi$ is a flavor $SU(2)$ doublet and a color $SU(N)$
$N$-column vector. The Lagrangian is diagonal in color, in contrast
with the full QCD Lagrangian which is diagonal in flavor. $\bfm{\tau}
=\{\tau_1,\tau_2,\tau_3\}$ are the three isospin Pauli matrices,
$\partial\slash = \gamma^\mu \partial_\mu$, and $m_0$ is the bare
quark mass (if $m_0 \neq 0$ the chiral symmetry is explicitly broken).
To obtain a Lagrangian that is quadratic in the fermionic fields we
introduce the scalar auxiliary field $\sigma$ and the three
pseudoscalar auxiliary fields $\bfm{\pi}=\{\pi_1, \pi_2, \pi_3\}$. Using
the functional identity \cite{Ebert-Volkov}
\begin{eqnarray}
& & \exp \left\{ i {G_1 \over 2} \int d^4x
\left[ (\Psibar\Psi)^2 + (\Psibar i \gamma_5 \bfm{\tau} \Psi)^2
\right] \right\} \sim \nonumber \\
& &\int[d\sigma d\bfm{\pi}]
\exp \left\{ i \int d^4x \left[-\Psibar
(\sigma+i \gamma_5 \bfm{\tau \cdot \pi})\Psi
-n_f \beta_1 (\sigma^2 + \bfm{\pi}^2)
\right] \right\}
\label{fun_identity}
\end{eqnarray}
the Lagrangian density becomes
\begin{eqnarray}
{\cal L} & = & \Psibar M \Psi - n_f \beta_1 (\sigma^2 + \bfm{\pi}^2)
\nonumber \\
M & = & i \partial\slash - m_0 - \sigma - i \gamma_5 \bfm{\tau \cdot
\pi} \fs
\label{Langrangian2}
\end{eqnarray}
Here $n_f=2$ is the number of flavors and $\beta_1 = {1\over 2 n_f G_1}$.
The fermionic fields can now be integrated and the resulting partition
function is
\begin{equation}
Z = \int[d\sigma d\bfm{\pi}] [\det M]^N
e^{-i n_f \beta_1 \int d^4x (\sigma^2 + \bfm{\pi}^2)} \fs
\label{PartFun1}
\end{equation}
Notice that there is no explicit kinetic energy term for the $\sigma$
and \bfm{\pi} fields. As we will show soon this term is part of $\det
M$.

In the general case the fermionic determinant has a phase that is
related to the the chiral anomaly and the Wess--Zumino term
\cite{Ebert-Reinhard}. In our work because the flavor space is
restricted to $SU(2)$ the phase is not present and we will write our
action in a symmetric fashion with regards to $M$ and $M^\dagger$.
Going to Euclidean space and appropriately discretizing the above
Lagrangian we obtain the model on the Euclidean Hypercubic lattice.
On the lattice, as it is well known, we have species doubling. The
doubling in the NJL model will be interpreted as a doubling of the
color degrees of freedom. To treat this problem we add to the
Lagrangian density an irrelevant operator (Wilson term) of the form
${a r \over 2} \Psibar \partial^2 \Psi$, ``$a$'' being the lattice
spacing and ``$r$'' a constant. We consider the $r=0$ case where no
effort is made to remove the doublers (naive fermions) and also the $r
\neq 0$ case where the doubler masses are raised to the cutoff (Wilson
fermions) and the chiral symmetry is explicitly broken.  With these
considerations and after appropriate scaling of the fields and
couplings, so that only dimensionless quantities appear, we obtain:
\begin{eqnarray}
Z &=& \int[d\Psi d\Psibar d\sigma d\bfm{\pi}] e^{-S} \nonumber \\
S &=& \sum_{x,y} \left\{ \sum_{i=1}^{N/2}
\left\{\Psibar^i_x M_{xy} \Psi^i_y +
\Psibar^{i+N/2}_x M^\dagger_{xy} \Psi^{i+N/2}_y
\right\} + n_f \beta_1 (\sigma^2_x + \bfm{\pi}^2_x) \delta_{xy}
\right\}
\nonumber \\
M_{xy} &=& {1\over 2} \sum_{\mu} \left[
(\gamma_\mu-r)\delta_{x+\mu,y} - (\gamma_\mu+r)\delta_{x-\mu,y}
\right]
+ (4 r + m_0 +
\sigma_x + i \gamma_5 \bfm{\pi_x\cdot\tau} )\delta_{xy}
\label{PartFun2}
\end{eqnarray}
with $\gamma_\mu$ hermitian. This partition function can be cast into
two different forms. The first can be studied using a large $N$
expansion and is given by:
\begin{eqnarray}
Z &=& \int[d\sigma d\bfm{\pi}] e^{-S_1} \nonumber \\
S_1 &=& N \left\{ -{1\over 2} {\rm Tr}(\log M) - {1\over 2}
{\rm Tr}(\log M^\dagger) + n_f \b_1 \sum_x (\sigma^2_x
+ \bfm{\pi}^2_x) \right\}
\label{PartFun4}
\end{eqnarray}
with $\b_1 = \beta_1 / N$ and the trace taken over space, spin,
flavor, and color. The second is appropriate for numerical simulations
and is given by:
\begin{eqnarray}
Z &=& \int[dX dX^\dagger d\sigma d\bfm{\pi}] e^{-S_2} \nonumber \\
S_2 &=& \sum_{x,y} \left\{ \sum_{i=1}^{N/2} \left\{ {X^i_x}^\dagger
(M^\dagger M)^{-1}_{xy} X^i_y \right\}+ n_f \beta_1 (\sigma^2_x +
\bfm{\pi}^2_y) \delta_{xy} \right\}
\label{PartFun3}
\end{eqnarray}
with $X$ being pseudofermionic fields.

\section{Large $N$}

We perform a standard large $N$ expansion with the action of eq.
\ref{PartFun4}. The large $N$ approximation will be reasonable for as
long as $\b_1$ is of order one. We assume a translation invariant
saddle and small fluctuations around it. For the case of naive
fermions using the chiral symmetry we rotate the fields so that the
saddle field configuration lies along the $\sigma$ direction.  For the
case of Wilson fermions the chiral symmetry is broken and the saddle
field configuration lies along the $\sigma$ direction.  As was shown
in \cite{Aoki}, the remaining parity-flavor symmetry can spontaneously
break generating a non zero vacuum expectation value for the pion
field.  However, in this work we will always stay in the parity-flavor
symmetric phase where the pion field has zero expectation value.  We
have:

\begin{equation}
\sigma(x) = \sigma_s + {\delta\sigma(x) \over \sqrt{N}},\ \ \
\bfm{\pi}(x) = {\delta\bfm{\pi}(x) \over \sqrt{N}}
\label{fields}
\end{equation}

In momentum space the inverse quark propagator at the saddle is:
\begin{equation}
\tilde{M}_s(p) = i \sum_\mu \gamma_\mu \sin p_\mu +
r (4 - \sum_\mu \cos p_\mu) + m_0 + \sigma_s \fs
\label{quark_prop}
\end{equation}
We identify
\begin{equation}
m_0 + \sigma_s = m_q
\label{quark_mass}
\end{equation}
as the quark mass. This definition is valid close to the continuum
limit where $m_q$ is small. It must be pointed out that $m_q$ is the
constituent quark mass since all high energy gluonic degrees of
freedom of QCD have been integrated out. In physical units the quark
mass will be taken to be equal to one third the proton mass.

We expand around the saddle (expansion in
${\delta\sigma(x) \over \sqrt{N}}$, ${\delta\bfm{\pi}(x) \over
\sqrt{N}}$). The zero order gives the effective potential
\begin{equation}
{V_{eff} \over N} = -2 n_f \int_{p} \log\left[g(p,m_q)\right] +
n_f \b_1 \sigma_s^2
\label{eff_pot}
\end{equation}
where $\int_{p} = \int_{p \in B} {d^4k \over (2 \pi)^4}$ for an
infinite lattice (B denotes the hypercubic lattice Brillouin zone) and
$\int_{p} = {1 \over L_x^3 L_t} \sum_{\{n_1,n_2,n_3,n_4\}}$ with
$n_1,n_2,n_3 \in [0, L_x-1]$, $n_4 \in [0, L_t-1]$ and $ p = ({2\pi
n_1 \over L_x}, {2\pi n_2 \over L_x}, {2\pi n_3 \over L_x}, {2\pi n_4
\over L_t})$ for a finite lattice with spatial extend $L_x$ and
temporal extent $L_t$. In the above equation $g(p,m_q)$ is defined as
\begin{eqnarray}
g(p,m_q) &=& \sum_\nu \sin^2 p_\nu + \left[r w(p) + m_q \right]^2
= p^2 + m_q^2 + O(p^2 m_q) \nonumber \\
w(p) &=& 4 - \sum_\mu \cos p_\mu = {p^2 \over 2} + O(p^4) \fs
\label{g-w}
\end{eqnarray}

The saddle point equations are obtained at the point where the linear
term vanishes.
\begin{equation}
\sigma_s {\b_1 \over 2} -
\int_p {\sigma_s + m_0 + r w(p) \over g(p,m_0+\sigma_s)} = 0
\label{sp_eq}
\end{equation}

The $\sigma$ and $\bfm{\pi}$ propagators $G_\sigma(q)$ and
$G_\pi(q)$ can be obtained from the
second order term of the expansion around the saddle
${1\over 2}\int_q \delta\sigma(q) G_\sigma^{-1} \delta\sigma(-q) +
{1\over 2}\int_q \delta\bfm{\pi}(q) G_\pi^{-1} \delta\bfm{\pi}(-q)$
with:
\begin{eqnarray}
G_{\pi/\sigma}^{-1}(q) &=&
4 n_f \left[ {\b_1 \over 2} - I_{\pi/\sigma}(q)\right]
\label{G} \\
I_{\pi/\sigma}(q) &=&\int_k {\sum_{\nu}
\sin(\kpnu) \sin(\kmnu)
+\!\!/\!\!-
\left[r w(\kp) + m_q \right] \left[r w(\km) + m_q \right]
\over
g(\kp, m_q) g(\km, m_q)}
\label{I}
\end{eqnarray}

We define the pion wave function renormalization constant $Z_\pi$ and
pion mass $m_\pi$ from:
\begin{equation}
\lim_{q \rightarrow 0}
G^{-1}_\pi(q) = Z_\pi^{-1} (q^2 + m_\pi^2) \fs
\label{Zpi-mpi_def}
\end{equation}
With this definition we find using eq. \ref{G} and
\ref{I}:
\begin{eqnarray}
Z_\pi^{-1} &=& {n_f \over 2} \int_k
{\sum_\mu\left[ \cos^2 k_\mu + r^2 \sin^2 k_\mu \right] \over g(k,
m_q)^2} \label{Zpi_full} \\
Z_\pi^{-1} m_\pi^2 &=& 4 n_f \left\{ {\b1 \over 2} - \int_k
{1 \over g(k, m_q)} \right\} \fs
\label{Zpi-mpi_full}
\end{eqnarray}
Notice that $Z_\pi$ is positive and therefore does not create local
stability problems (the propagator is positive). Also notice that with
this definition of the pion mass, $m_\pi$ will be the true pole of
$G_\pi$ only if $m_\pi$ is small. For the cases we are interested in
this will always be true. Unfortunately we can not define the $\sigma$
mass $m_\sigma$ and width $\gamma_\sigma$ in a similar way since
$m_\sigma$ will not be small in general. The proper
definition in this case is:
\begin{equation}
G^{-1}_\sigma(q=\{i
m_\sigma + {\gamma_\sigma \over 2},0,0,0\}) = 0
\label{msig_gamsig_def}
\end{equation}
and we can not obtain closed form expressions for
$m_\sigma$ and $\gamma_\sigma$.

On a finite lattice the finite momentum sums can be calculated with
the aid of a computer and the above quantities except $\gamma_\sigma$
can be exactly determined. This will be done in Section 4.1 when we
will compare the numerical results obtained on a finite lattice with
the large $N$ predictions for the same lattice size.  For an infinite
lattice the momentum integrals can be evaluated numerically but this
will not serve our purposes since we are interested in obtaining
analytical expressions. For this reason we resort into performing
asymptotic expansions in small $m_q$, $m_\pi$, $m_\sigma$, and $q$,
considering the logarithms as being of order zero. The asymptotic
expansions will be given separately for the $r =0$ and $r \neq 0$
cases in the next two sections.

\subsection{Naive fermions at infinite volume, $r=0$}

In this case we have species doubling. The doubling is interpreted as
a ``doubling" of the color degrees of freedom. That this
interpretation is appropriate can be seen from eq. \ref{eff_pot}. For
$r=0$ the function $g(p,m_q)$ of eq. \ref{g-w} is periodic with period
$\pi$ and not $2 \pi$. This means that the integral in eq.
\ref{eff_pot} splits into $16$ equal pieces and therefore the
effective potential will be made up from $16 N$ copies. Therefore
we set the number of colors $N_c$ to:
\begin{equation}
N_c = 16 N \fs
\label{Nc_r0}
\end{equation}

Some of the typical integrals we will encounter are:
\begin{eqnarray}
J_{n}(m_q) &=& \int_k {1 \over g(k,m_q)^n}
\nonumber \\
I_0(q,m_q) &=& \int_k {1 \over g(\kp, m_q) g(\km, m_q)} \fs
\label{Jn-I0}
\end{eqnarray}
Simple trigonometric relations relate these integrals with the
corresponding integrals that arise for Bose particles. We get
\begin{eqnarray}
J_n(m) &=& 16 \left\{ 4^{n-2} J_n^{B}(2m) \right\}\nonumber \\
I_0(q,m) &=& 16 I_0^{B}(2q,2m)
\label{Fermi-Bose}
\end{eqnarray}
where $J_n^{B}$ and $I_0^{B}$ are defined as in eq. \ref{Jn-I0} but
with $g(k,m)=2 \sum_\nu \left[ 1 - \cos(k_\nu) \right] + m^2 $ the
inverse propagator of a Bosonic particle. The asymptotic expansions of
$J_{1}^{B}$ and $J_{2}^{B}$ have been evaluated to leading order in
$m_q$ with very accurately determined coefficients in
\cite{Luscher-Weisz_1} Appendix B. The leading order term of $I_0^{B}$
is universal except for a lattice constant. For example, it has been
evaluated for the $F_4$ lattice in \cite{BBHN1}. The lattice constant
needed also appears in the leading order term of the asymptotic
expansion of $J_{2}^{B}$ and for the hypercubic lattice can be taken
from \cite{Luscher-Weisz_1}. For the convenience of the reader we
present these results below:
\begin{eqnarray}
J_1^{B}(m) &=&  r_0 + m^2(r_1 + s_1 \log m^2) + O(4) \nonumber \\
J_{n+1}^{B}(m) &=& -{1 \over n} {\partial \over \partial m^2}
J_{n}^{B}(m) \nonumber \\
I_0^{B}(q,m) &=& -s_1 \int_{0}^{1} ds \log[m^2 + s(1-s)q^2]
- (r_1 + s_1) + O(2)
\label{Bose}
\end{eqnarray}
with
\begin{eqnarray}
r_0 &=& 0.154\ 933\ 390 \nonumber \\
r_1 &=& -0.030\ 345\ 755 \nonumber \\
s_1 &=& {1 \over 16 \pi^2} \fs
\label{Bose_const}
\end{eqnarray}

Because $r=0$, if we set $m_0=0$ the chiral symmetry is not
explicitly broken. In that case there are two solutions to the saddle
point equation \ref{sp_eq}, namely $\sigma_s = 0$ and $\sigma_s \neq
0$. For $\b_1 > \b_{1_c}$, the dominating saddle is $\sigma_s = 0$ and
we are in a chirally symmetric phase with massive $\sigma$ and
$\bfm{\pi}$ fields and massless quarks $m_q = \sigma_s = 0$. For $\b_1
< \b_{1_c}$ the dominating saddle is $\sigma_s \ne 0$ and we are in a
phase with spontaneously broken chiral symmetry. The pions are the
Goldstone bosons and become massless. The sigma is the massive mode
and the quarks acquire a dynamically generated mass $m_q = \sigma_s
\ne 0$. The critical value of $\b_1$ is:
\begin{equation}
\b_{1_c} = 2 J_1(0) = 8 r_0 \fs
\label{beta1_c}
\end{equation}
All of our analysis will be done in the broken phase.

The asymptotic expansion of $G^{-1}_\pi(q)$ is:
\begin{equation}
G_{\pi}^{-1}(q) =
Z_\pi^{-1} m_\pi^2 + 4 n_f q^2 \left\{ {1 \over 2} I_0(q,m_q)
-{1\over 8} J_1(0) \right\} + O(4)
\label{Gpi_as}
\end{equation}
where only the leading order term of $Z_\pi^{-1} m_{\pi}^2$ and $I_0$
is to be kept. The coefficient of the $q^2$ term will be equal to
$Z_\pi$ for $q \rightarrow 0$. The pion mass mass should satisfy
$G^{-1}_\pi(q=\{i m_\pi,0,0,0\}) = 0$. To this order in the asymptotic
expansion this will be true only if the coefficient of the $q^2$ term
for $q=\{i m_\pi,0,0,0\}$ is very close to $Z_\pi$. From the
asymptotic expansion of $I_0(q,m_q)$ (eq. \ref{Fermi-Bose},
\ref{Bose}) we see that this will happen for $m_\pi^2 s (1-s) \ll
m_q^2$ where $s \in [0,1]$. Therefore for this definition of $m_\pi$
to be valid we must not only demand that $m_\pi$ is small, as mentioned
in the previous section, but also that:
\begin{equation}
m_\pi^2 \ll 4 m_q^2 \fs
\label{mpi_constrain}
\end{equation}
For all the cases we will be interested in this will be satisfied. For
example with $M_q=310 \MeV$ and $M_\pi=140 \MeV$ we get ${M_\pi^2
\over 4 M_q^2} = 0.054$.

Equation \ref{Zpi_full} can be rewritten in the form:
\begin{equation}
Z_\pi^{-1} = {n_f \over 2} \left[ (4+m_q^2) J_2(m_q) -
J_1(m_q)\right] \fs
\label{Zpi_full_r0}
\end{equation}
Using eq. \ref{Fermi-Bose}, \ref{Bose} and \ref{Zpi_full_r0} we find
the leading order term of the asymptotic expansion of $Z_\pi$:
\begin{eqnarray}
Z_\pi^{-1} &=& 16 [ z_0 - z_1 \log m_q^2 ] \nonumber \\
z_0 &=& -2 n_f ({r_0 \over 16} + r_1 + s_1 + s_1 \log 4),\  {\rm for}
\ \ n_f=2 \ \ \ z_0 = 0.022\ 204\ 130 \nonumber \\
z_1 &=& 2 n_f s_1 \fs
\label{Zpi_as_r0}
\end{eqnarray}
As mentioned in the previous section $Z_\pi$ is always positive and
does not create local stability problems.  From the above equation we
find that $Z_\pi$ will become negative if $m_q > \exp{z_0 \over 2 z_1}
\simeq 1.55$. This provides us with a point above which the small
$m_q$ approximation is certainly not valid. In our analysis we will
never need to be close to this point. However, the above equation has a
much more important consequence. It can be shown \cite{Ebert-Volkov}
that the pion decay constant $f_\pi$ is related to $m_q$ and $Z_\pi$
via:
\begin{equation}
f_\pi^2 = N m_q^2 Z^{-1}_\pi \fs
\label{fpi1}
\end{equation}
Using eq. \ref{fpi1}, \ref{Zpi_as_r0} and \ref{Nc_r0} we get:
\begin{equation}
{f_\pi^2 \over m_q^2} = N_c [ z_0 - z_1 \log m_q^2 ] \fs
\label{fpi2}
\end{equation}
Then with $N_c=3$, $n_f=2$, ${f_\pi \over m_q} = {F_\pi \over M_q} =
{93 \MeV \over 310 \MeV} = 0.3$ and with $m_q = M_q a$, $\Lambda =
{\pi \over a}$ we find $\Lambda = 1150 \MeV$. This is consistent with
the expectation that the cutoff of the theory should be $\ltapprox
1550 \MeV$ (the mass of the lightest glueball \cite{UKQCD}).

Although, as mentioned in the introduction, it is not the aim of this
work to study the possibility of the Higgs being a top quark
condensate, we point out that similar considerations as in the
previous paragraph can be used to estimate the cutoff of such a
theory. With $N_c=3$, $n_f=1$, $F_\pi=246 \GeV$, $M_q=170 \GeV$ we
obtain $\Lambda = 3 \times 10^{14} \GeV$. This is consistent with the
much more complete and detailed analysis of \cite{Bardeen}.

Equation \ref{Zpi-mpi_full} can be rewritten in the form:
\begin{equation}
m_\pi^2 = 4 n_f Z_\pi {m_0 \over m_q - m_0} J_1(m_q)
\label{mpi_full_r0}
\end{equation}
where the saddle point equation \ref{sp_eq} was used to eliminate
$\b_1$. Using eq. \ref{mpi_full_r0}, \ref{Bose} and \ref{Fermi-Bose},
we find the leading order term of the asymptotic expansion of $m_\pi$:
\begin{equation}
m_\pi^2 = {m_0 \over (m_q-m_0)} {n_f r_0 \over [z_0 - z_1 \log m_q^2]}
+ O(2) \fs
\label{mpi_as_r0}
\end{equation}
Notice that when the chiral symmetry is not explicitly broken, namely
$m_0 = 0$, then in the broken phase where $m_q-m_0 = \sigma_s \neq 0$,
$m_\pi = 0$ as it should.

Next we concentrate on $m_\sigma$ and $\gamma_\sigma$.
Using the saddle point equation \ref{sp_eq} to eliminate $\b_1$
and eq. \ref{Zpi_as_r0} the leading order term of the asymptotic
expansion of the inverse sigma propagator of eq. \ref{G} is:
\begin{equation}
{G_{\sigma}^{-1}(q) \over 16}=
{M_\pi^2 \over F_\pi^2} N_c [z_0 - z_1 \log m_q^2]^2 m_q^2
+ 4 n_f \left\{ {1 \over 2} (q^2 + 4 m_q^2) I_0^{B}(2q,2m_q)
-{r_0\over 32} q^2 \right\} + O(4)
\label{Gsig_as}
\end{equation}
where only the leading order term of $I^B_0$ given in eq. \ref{Bose}
is to be kept. The leading order term of $I_0^B$ contains a $\log(-1)$
that we take to be equal to $+ i \pi$ (taking it to be equal to $-i
\pi$ results to a negative width). Using the above equation and the
definition \ref{msig_gamsig_def} we can calculate $m_\sigma$ and
$\gamma_\sigma$. Because we can not get a closed form expression we
solve numerically and the result is shown in Figure 1.

{}From Figure 1 we find that if we set the quark mass to one third the
proton mass $M_q = 310 \MeV$ then $M_\sigma = 726 \MeV$,
$\Gamma_\sigma=135 \MeV$, and $\Lambda = \pi / a = 1150 \MeV$.
$M_\sigma$ is consistent with phenomenological expectations and as
mentioned earlier $\Lambda$ is consistent with the expectation that
the cutoff should be close and below the mass of the lightest glueball
($1550 \MeV$). The width however is underestimated. The reason is
traced to the fact that to leading order in large $N$ the width
receives contributions only from the quark bubble and not from the
pion bubble because the pion bubble is of order $1/N$. Because the
phase space available for the $\sigma$ to decay to two quarks is much
smaller than the phase space to decay to two pions, the pion loop
contribution, although of order $1/ N$, is probably more important
than the quark loop contribution.

The above result can also be used to make an interesting observation.
If the Higgs sector is the low energy effective field theory of a NJL
model with exactly the same parameters as the low energy QCD except
for $M_\pi=0$ and $F_\pi=246 \GeV$, then we find $M_\sigma = 1915\GeV$.
This corresponds to $m_\sigma = 2$ where one would expect very large
deviations from the low energy behavior of scattering cross sections.
Although we have not calculated these deviations the value of the
width serves as an indication of their size. In a way, departure from
low energy behavior will be signaled by an increasing width of the
$\sigma$ to two quark decay. At $m_\sigma \approx 2$ the width is
already fairly large.

\figure{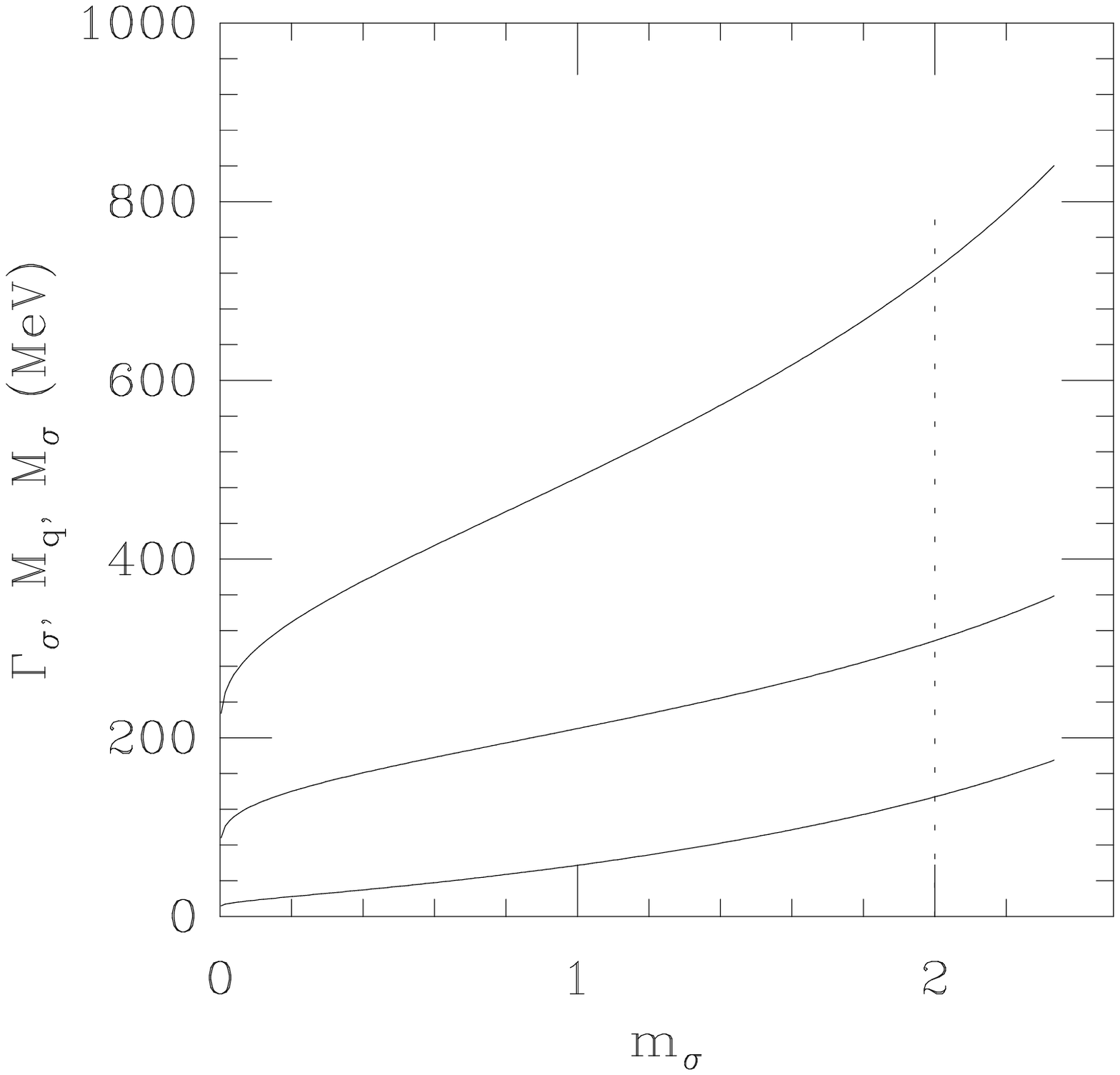}{\one}{\figsizeA}

As a final comment notice that close to the continuum limit and by
setting the renormalization point of the integral $I_0^{B}$ in eq.
\ref{Gsig_as} at $q=0$  we obtain
\begin{eqnarray}
m_\sigma^2 &=&
m_\pi^2 + 4 m_q^2 \nonumber \\ \gamma_\sigma &=& 0
\label{cont}
\end{eqnarray}
where we have used the definition of eq. \ref{msig_gamsig_def}.  This
is the result obtained in the literature with continuum type
regularizations (see for example \cite{Ebert-Volkov}). This result,
because of the off--shell renormalization point, is only approximate
and also neglects the contribution of the quarks to $\gamma_\sigma$.

\subsection{Wilson fermions at infinite volume, $r \neq 0$}

In this case the doublers have been removed by raising their masses to
the cutoff. The chiral symmetry has been explicitly broken by the
Wilson term and the pions are massive. The number of colors is:
\begin{equation}
N_c = N \fs
\label{Nc_r}
\end{equation}

Some of the typical integrals we will encounter are:
\begin{eqnarray}
J_{n,m}(m_q) &=& \int_k {w(k)^n \over g(k,m_q)^m}
\nonumber \\
I_0(q,m_q) &=& \int_k {1 \over g(\kp, m_q) g(\km, m_q)} \fs
\label{J-I0_r}
\end{eqnarray}
Unfortunately there are no simple trigonometric relations that relate
these integrals with the corresponding integrals that arise for Bose
particles as in the $r=0$ case. We will need the asymptotic
expansions of $J_{0,1}$, $J_{1,1}$ to order $m_q^2$ and the leading
order term of $J_{0,2}$ . We have calculated them and the result is
given in Appendix A.  We will also need the leading order asymptotic
expansion of the $I_0$ integral. This, up to the lattice constant
$r_1$ that needs to be calculated and is given in Appendix A, is the
same as the leading order term of $I^B_0$ in eq. \ref{Bose}.  Some of
the lattice constants that we will need can be parametrized as:
\begin{equation}
a_{n,m} = J_{n,m}(0)\ \ \ \ \  m-n \leq 1 \fs
\label{anm}
\end{equation}
In our analysis only few of them appear and we have calculated them
for $r=1$ and $r=0.1$. They are given in Appendix A together with the
constants
\begin{eqnarray}
r_1 &=& \int_k \left\{{1 \over g(k,0)^2} -
{\Theta(l^2 - k^2) \over k^4} \right\} + {1 \over 16 \pi^2}\log(l^2)
\nonumber \\
s_1 &=& {1 \over 16 \pi^2} \fs
\label{const_r}
\end{eqnarray}
where $\Theta$ is the step function.

The asymptotic expansion of $G^{-1}_\pi(q)$ is:
\begin{equation}
G_{\pi}^{-1}(q) =
Z_\pi^{-1} m_\pi^2 + 4 n_f q^2 \left\{ {1 \over 2} I_0(q,m_q)
+ {(r^2-1)\over 8} a_{0,1} - {r^2(r^2-1)\over 8} a_{2,2} \right\}
+ O(3)
\label{Gpi_as_r}
\end{equation}
where only the leading order term of $Z_\pi^{-1} m_{\pi}^2$ and $I_0$
is to be kept. The same restrictions on $m_\pi$ as in eq.
\ref{mpi_constrain} are needed.

The leading order term of the asymptotic expansion of $Z_\pi$ is:
\begin{eqnarray}
Z_\pi^{-1} &=& [ z_0 - z_1 \log m_q^2 ] \nonumber \\
z_0 &=& -2 n_f \left[ r_1 + s_1 + {1 - r^2 \over 4} a_{0,1}
- {r^2 (1 - r^2) \over 4} a_{2,2} \right],\ {\rm for}\ \
n_f=2\ ,r = 1,\ \ \  z_0 = 0.0223 \nonumber \\
z_1 &=& 2 n_f s_1 \fs
\label{Zpi_as_r}
\end{eqnarray}
Notice that for $r=1$, $z_0$ has a value that is very close to the one
of the $r=0$ case. Because of that and since $z_1$ is universal all
the discussion relating to eq. \ref{Zpi_as_r0}, \ref{fpi1} and
\ref{fpi2} is also valid here.

Equation \ref{Zpi-mpi_full} can be rewritten in the form:
\begin{equation}
m_\pi^2 = 4 n_f Z_\pi {1 \over m_q - m_0}
\left[ m_0 J_{0,1}(m_q) + r J_{1,1}(m_q) \right]
\label{mpi_full_r}
\end{equation}
where the saddle point equation \ref{sp_eq} was used to eliminate
$\b_1$. The leading order term of the asymptotic expansion of $m_\pi$
is:
\begin{equation}
m_\pi^2 =
{4 n_f \over (m_q-m_0)}
{m_0 a_{0,1} + r a_{1,1} \over [ z_0 - z_1 \log m_q^2 ]}
+ O(2) \fs
\label{mpi_as_r}
\end{equation}

We are now in a position to investigate the approach to the continuum
chiral limit. The theory has two adjustable bare parameters,
$\beta_1$ and $m_0$. The parameter $m_0$ is related to the more familiar
hoping parameter $\kappa$ (often used in QCD) by:
\begin{equation}
\kappa = {1 \over 8 r + 2 m_0} \fs
\label{kappa}
\end{equation}
The two bare parameters control $m_q$ and $m_\pi$ through equations
\ref{sp_eq} and \ref{mpi_full_r}. The following statements can be made:

\noindent{\bf{A}}
The $m_q = 0$ line where the continuum limit is retrieved is defined
by:
\begin{equation}
\beta_1 + { 2 N_c r a_{1,1} \over m_0} = 0
\label{mq0_line}
\end{equation}

\noindent{\bf{B}}
For any $m_q$, $m_0$ can be adjusted so that $m_\pi =0$. The $m_\pi=0$
line is given by:
\begin{equation}
m_0 J_{0,1}(m_q) + r J_{1,1}(m_q) = 0
\label{mpi0_line}
\end{equation}
As was shown in \cite{Aoki} this line separates the flavor-parity
symmetric phase from the flavor-parity broken phase. In this work we
always stay in the symmetric phase.

\noindent{\bf{C}}
The continuum chiral limit is obtained when both $m_q$ and $m_\pi$ go
to zero. This point is:
\begin{equation}
m_{0_{chiral}} = -r {a_{1,1} \over a_{0,1}} ,\ \ \ \
\beta_{1_{chiral}} = - {2 N_c r a_{1,1} \over m_0}
\label{mq0_mpi0_line}
\end{equation}
and all physics should be extracted at the vicinity of this point.

Using the asymptotic expansions of eq. \ref{sp_eq} and
\ref{mpi_full_r} to order $m_q^2$, we plot for small $m_q$ and $m_\pi$
the constant $m_q$ and constant $m_\pi$ lines for $r=1$ in Figure 2a
and for $r=0.1$ in Figure 2b. The two Figures are qualitatively the
same with the $m_q=0$, $m_\pi=0$ point shifted toward larger $\kappa$
and $\beta_1$. In fact as $r \rightarrow 0$ it can be easily shown
that $\beta_1 \rightarrow N \b_{1_c}$, where $\b_{1_c}$ is the
critical value of the $r=0$ case (see eq. \ref{beta1_c}), and $\kappa
\rightarrow \infty$.  From eq. \ref{mpi_as_r}
we find that as $m_q \rightarrow 0$ also $m_\pi \sim 1 / \log m_q^2
\rightarrow 0$.  This implies that the $m_q=0$ line is also an $m_\pi =
0 $ line. This behavior is apparent in Figure 2a and 2b. However, this
is an artifact of our approximation in defining $m_\pi$. As mentioned
earlier (see the comment after eq. \ref{Gpi_as_r}) our definition of
$m_\pi$ will be valid only if $m_\pi^2 \ll 4 m_q^2$ which can
not be satisfied on the $m_q =0$ line except on the one point where it
intersects the $m_\pi=0$ line. The constant $m_\pi$ lines are
therefore valid only in the regions where $m_\pi^2 \ll 4 m_q^2$. To
get the full constant $m_\pi$ lines  a more detailed analysis would
be necessary

\figure{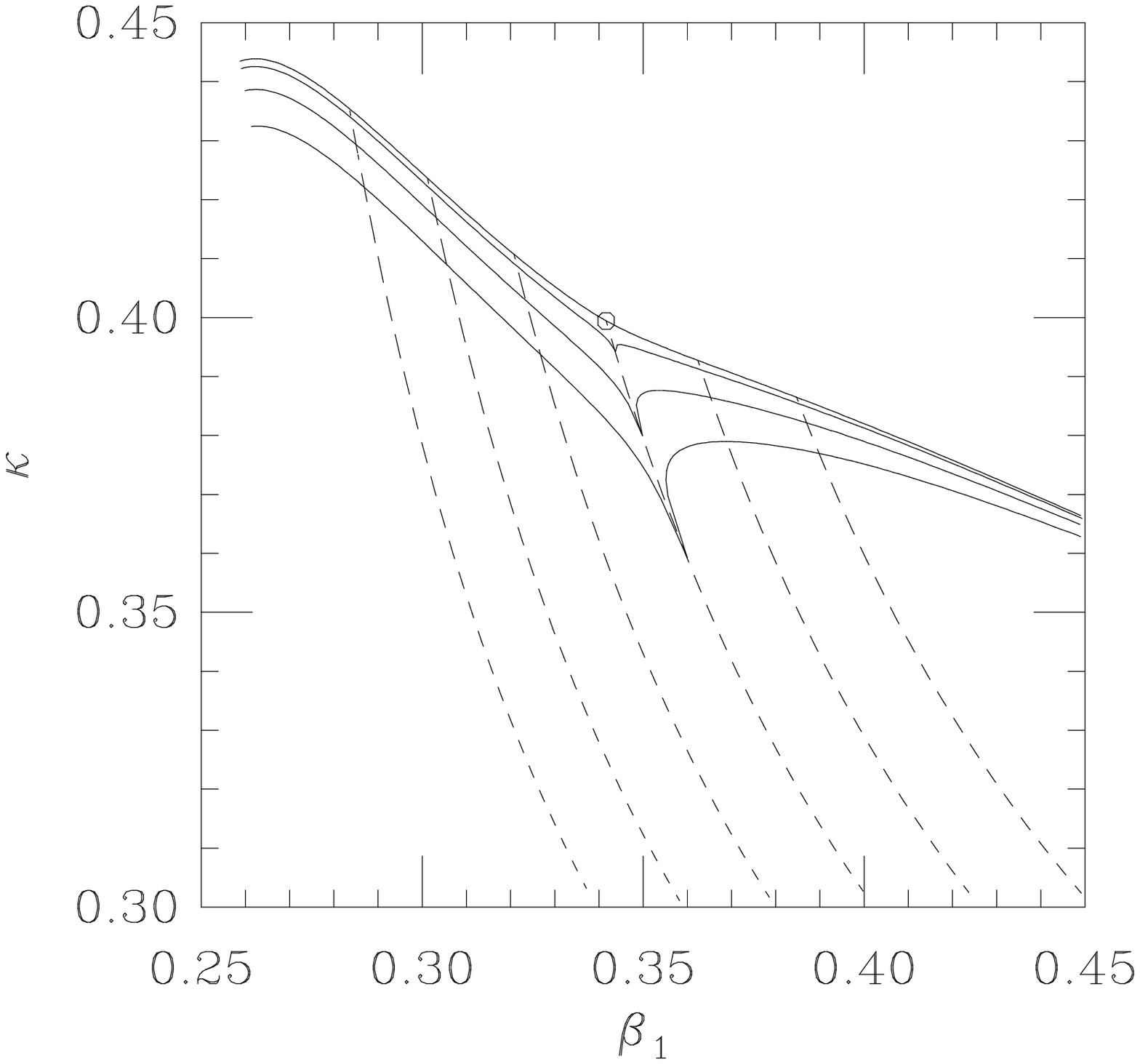}{\twoA}{\figsizeA}

{}From these Figures we see that if for a fixed $\beta_1$ we were to
change $\kappa$ from smaller to larger values (as is often done in QCD
with dynamical Wilson fermions) then if $\beta_1 < \beta_{1_{chiral}}$
we would reach the $m_\pi =0$ limit before we reach the continuum
limit. On the other hand if $\beta_1 > \beta_{1_{chiral}}$ we would
reach the continuum limit before we reach the $m_\pi =0$ limit. As
mentioned in ``{\bf C}" above, there is only one point in the
$\beta_1$, $\kappa$ plane where we can obtain a continuum chiral
limit.

Using the saddle point equation \ref{sp_eq} the chiral condensate is
found to be:
\begin{equation}
{\langle \Psibar \Psi \rangle \over n_f} = -2 \b_1 (m_q -m_0)
= -2 \b_1 \sigma_s
\label{psibar_psi}
\end{equation}
\figure{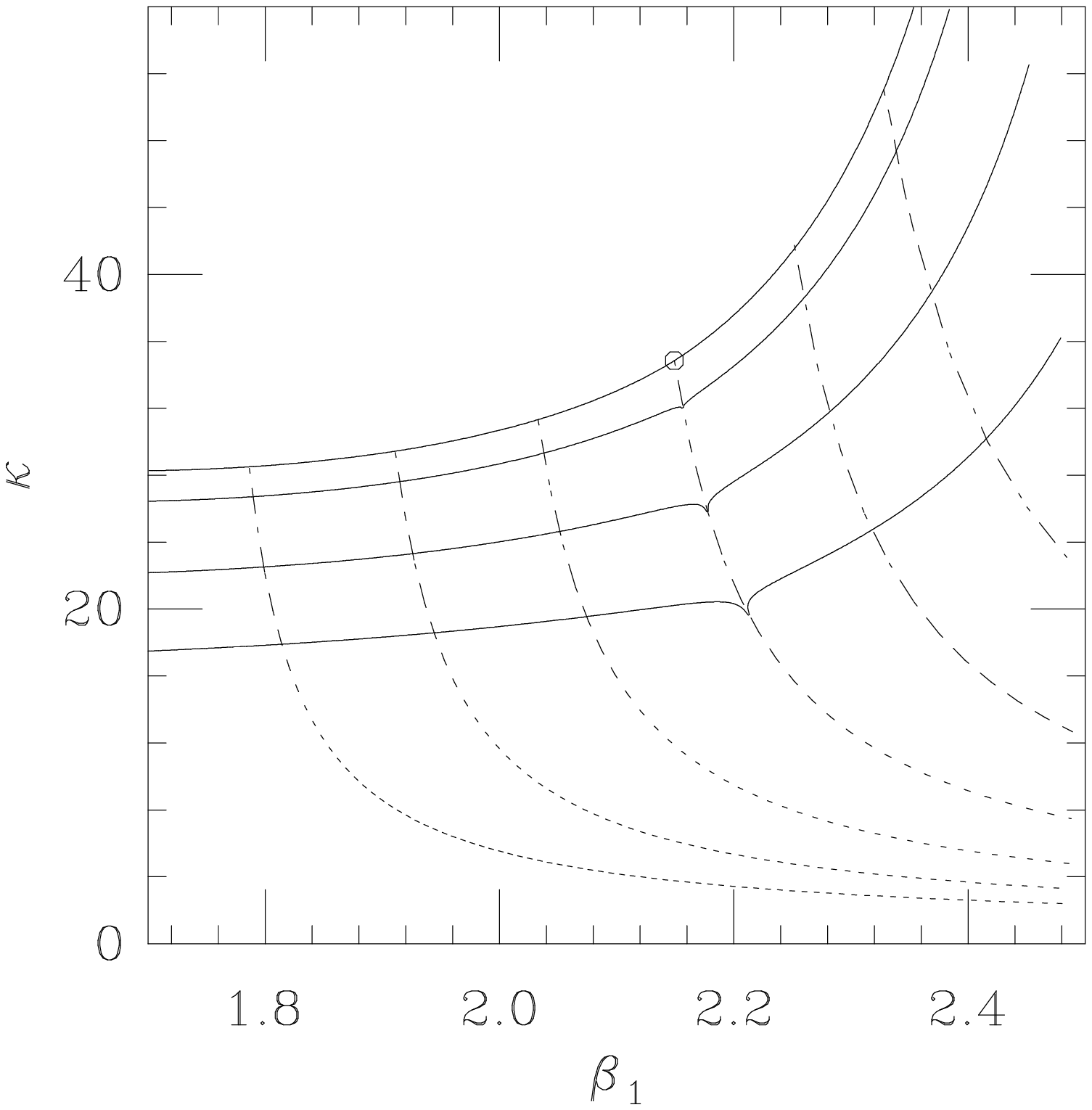}{\twoB}{\figsizeA}

\noindent
and it is not zero when $m_q=0$ and/or $m_\pi=0$. Therefore it can not
serve as an order parameter for Wilson fermions. On the other hand
when $r=0$ and $m_0=0$ $\langle \Psibar \Psi \rangle $ is an order
parameter because $\sigma_s$ is an order parameter.

Next we concentrate on $m_\sigma$ and $\gamma_\sigma$. The leading
order term of the asymptotic expansion of the inverse sigma propagator
of eq. \ref{G} is:
\begin{eqnarray}
G_{\sigma}^{-1}(q)
&=& {M_\pi^2 \over F_\pi^2} N_c [z_0 - z_1 \log m_q^2]^2 m_q^2
\nonumber \\
&+& 4 n_f\left\{
{1 \over 2} (q^2 + 4 m_q^2) I_0(q,m_q)
+ q^2 {(r^2-1)\over 8} a_{0,1}
+ R_0 + m_q R_1 + m_q^2 R_2 + q^2 R_3 \right\} \nonumber \\
&+& O(3)
\label{Gsigma_as_r}
\end{eqnarray}
where only the leading order term of $I_0$ is to be kept. $R_0$,
$R_1$, $R_2$, and $R_3$ are lattice constants defined below.
\begin{eqnarray}
R_0 &=& 2 r^2 a_{2,2} \nonumber \\
R_1 &=& 4 r a_{1,2} - 8 r^3 a_{3,3} \nonumber \\
R_2 &=& 24 r^4 a_{4,4} - 20 r^2 a_{2,3} \nonumber \\
R_3 &=& -{r^2-1 \over 16} R_0 \nonumber \\
& & -{r^2 \over 4}
\int_{k}{1 \over g^4(k,0)} \sum_{\mu}\left\{
\sin^2 k_\mu \left[g(k,0) -
2r^2w^2(k)-2w(k)\cos(k_\mu)\right]^2\right\}
\label{lat_consts}
\end{eqnarray}
$R_0$, $R_1$, $R_2$ can be calculated from the $a_{n,m}$'s.  $R_3$ has
to be calculated separately and is given in Appendix A.  Notice that
$R_0$, $R_1$, $R_2$ and $R_3$ go to zero for vanishing $r$.

It is immediately apparent that the $R_0$ and $m_q R_1$ terms do not
scale appropriately. As a result
\begin{equation}
m_\sigma^2 \sim R_0
\label{m_sig}
\end{equation}
and therefore $M_\sigma$ is of order cutoff.  For the NJL model at
large $N$ we can trace the reason for this phenomenon and offer an
exact answer. Such a phenomenon may also be responsible for the
difficulty in observing a $\sigma$ particle in numerical simulations
of QCD with dynamical Wilson fermions.

First we realize that although the Wilson term has raised the doubler
masses to the cutoff, the doublers have not disappeared and they can
possibly contribute through vacuum polarization effects. We must then
try to separate their contribution. Towards this end consider the
defining equation of the $\sigma$ propagator eq. \ref{G}.  The $\b_1 /
2$ term contributes to the ${M_\pi^2 \over F_\pi^2} m_q^2$ term of eq.
\ref{Gsigma_as_r}. This term scales ``correctly" and it will not
concern us. We focus on $I_\sigma$ of eq. \ref{I}. $I_\sigma(q) \sim
{\rm Re}\ {\rm Tr} \int_k \tilde{M}_s^{-1}(\km) \tilde{M}_s^{-1}(\kp)$
where $\tilde{M}_s^(p)$ is the quark propagator at the saddle point
and is given in eq.  \ref{quark_prop}. This integral corresponds to a
bubble integral with two external legs $\sigma(q)$ and $\sigma(-q)$
and two quarks flowing in the bubble with momenta $k-q/2$ and $k+
q/2$. Now, each species has momentum that belongs in a section $B_i,\
i = 1,2,\ldots 16$ of the Brillouin zone $B$ with extent $\pi/2$ and
$-\pi/2$ from the origin in each direction (see the first two columns
of Table 1). For small $q$ both quarks will have momentum around $k$
and we can then separate the contributions of each species by
splitting the integral into the $16$ regions $B_i$.  This way we can
isolate the contribution to each term in $I_\sigma$ from the
propagating quark and the 15 doublers. We will only discuss the
contributions to the $R_0$ term since $m_\sigma^2 \sim R_0$, but
similar arguments hold for the $R_1$, $R_2$ and $R_3$ terms.

{}From the definitions in \ref{lat_consts}, \ref{anm}, \ref{J-I0_r} and
splitting the integral into $16$ regions as described in the previous
paragraph we obtain:
\begin{equation}
R_0 = 2 r^2 \sum_{i=1}^{16} \int_{B_i} {w(k)^2 \over g_r(k,0)^2}
\label{R0_1}
\end{equation}
where we have explicitly denoted the $r$ dependence of $g(k,0)$.  It
is immediately apparent that the contribution, $R_0^{B_1}$, form the
$B_1$ region where the propagating quark lives is not zero. Therefore
we can not attribute the whole $R_0$ term to the doublers.

The contribution of the propagating quark comes from the high momentum
section of the $B_1$ region. To see this consider a spherical section
${\cal M} \sim m_q$ centered around the origin $(0,0,0,0)$ of $B_1$.
We have:
\begin{equation}
R_0^{B_1} = 2 r^2 \int_{\cal M} {w(k)^2 \over g_r(k,0)^2} +
2 r^2 \int_{B_1 \cap \cal M} {w(k)^2 \over g_r(k,0)^2} \fs
\label{R0_2}
\end{equation}
Since $W(k)\sim k^2$ and $g_r(k,0) \sim k^2$ (see eq. \ref{g-w}) we
find $\int_{\cal M} {w(k)^2 \over g_r(k,0)^2} \sim m_q^4$ and therefore
\begin{equation}
R_0^{B_1} = 2 r^2 \int_{B_1 \cap \cal M} {w(k)^2 \over g_r(k,0)^2}
+ O(m_q^4) \fs
\label{R0_3}
\end{equation}
The Wilson term has not only raised the masses of the doublers but has
also changed the high frequency behavior of the propagating quark.
In a bubble integral this change is visible.

We could contrive a Wilson term that will raise the doubler masses but
not change the high frequency behavior of the propagating quark. This
can be done by introducing a momentum dependent $r$ such that:
\begin{eqnarray}
r(p) &=& 0 \ \ {\rm if} \ p \in B_1 \nonumber \\
r(p) &=& r \ \ {\rm if} \ p \in B_2,\ B_3,\  \ldots B_{16} \fs
\label{rp}
\end{eqnarray}
Then
\begin{equation}
R_0 = 2 \int_{B} r(k)^2 {w(k)^2 \over g_r(k,0)^2} =
2 r^2 \sum_{i=2}^{16} \int_{B_i} {w(k)^2 \over g_r(k,0)^2}
\label{R0_4}
\end{equation}
and the $\sigma$ mass $m_\sigma \sim R_0$ will be composed
entirely from contributions due to the doublers.

The above discussion can become more transparent for $r \ll 1$.  The
masses of the $16$ species are the roots of $\sum_{\nu} \sin^2 k_\nu +
r^2 (4 - \sum_\nu \cos k_\nu)^2$. For small $r$ they are given in
Table 1.

\if \PRD N

\begin{table}
\begin{center}
\begin{tabular}{||c|c|c||}   \hline
``Brillouin zone'' & origin  & species mass $+ O(r^2)$  \\ \hline\hline
$B_{1 }$ & $   0 \    0 \    0 \    0$ & $m_{1 }= 0  $   \\ \hline
$B_{2 }$ & $ \pi \    0 \    0 \    0$ & $m_{2 }= 2r $   \\ \hline
$B_{3 }$ & $   0 \   \pi\    0 \    0$ & $m_{3 }= 2r $   \\ \hline
$B_{4 }$ & $   0 \    0 \   \pi\    0$ & $m_{4 }= 2r $   \\ \hline
$B_{5 }$ & $   0 \    0 \    0 \  \pi$ & $m_{5 }= 2r $   \\ \hline
$B_{6 }$ & $ \pi \  \pi \    0 \    0$ & $m_{6 }= 4r $   \\ \hline
$B_{7 }$ & $ \pi \    0 \  \pi \    0$ & $m_{7 }= 4r $   \\ \hline
$B_{8 }$ & $ \pi \    0 \    0 \  \pi$ & $m_{8 }= 4r $   \\ \hline
$B_{9 }$ & $  0  \  \pi \  \pi \    0$ & $m_{9 }= 4r $   \\ \hline
$B_{10}$ & $  0  \  \pi \    0 \  \pi$ & $m_{10}= 4r $   \\ \hline
$B_{11}$ & $  0  \    0 \   pi \  \pi$ & $m_{11}= 4r $   \\ \hline
$B_{12}$ & $  0  \  \pi \  \pi \  \pi$ & $m_{12}= 6r $   \\ \hline
$B_{13}$ & $ \pi \    0 \  \pi \  \pi$ & $m_{13}= 6r $   \\ \hline
$B_{14}$ & $ \pi \  \pi \    0 \  \pi$ & $m_{14}= 6r $   \\ \hline
$B_{15}$ & $ \pi \  \pi \  \pi \    0$ & $m_{15}= 6r $   \\ \hline
$B_{16}$ & $ \pi \  \pi \  \pi \  \pi$ & $m_{16}= 8r $   \\ \hline
\end{tabular}
\end{center}
\end{table}

\fi

We calculate the leading order term of the asymptotic expansion
of $R_0$ in $m_i$ and find:
\begin{equation}
R_0 = 2 r^2 \int_{B_1} {w(k)^2 \over g_{r=0}(k,0)^2} -
2 \sum_{i=1}^{16} m_i^2 \left[ r_1 + s_1 + s_1 \log m_i^2 \right]
+ O(4)
\label{R0_5}
\end{equation}
with $r_1$ and $s_1$ given in eq. \ref{const_r}. The first term comes
because of the change of the high frequency behavior of the propagator
of each species. The second term clearly shows that $m_\sigma^2$ can be
written as a sum over the doubler masses squared.

\section{Numerical and large $N$ work on finite lattices}

We simulate the action in eq. \ref{PartFun3} for $N=2$ using a
standard Hybrid Monte Carlo (HMC) algorithm \cite{DKPR}. We use the
conjugate gradient (CG) algorithm without preconditioning to invert
the matrix $M^{\dagger}M$ and the leap--frog algorithm to integrate
the equations of motion. We measure expectation values of operators
that involve the $\sigma$ and $\bfm{\pi}$ fields as well as operators
that involve the pseudofermionic fields or traces of appropriate
combinations of the matrix $M$.

The reader will realize that the matrix $M$ of eq.  \ref{PartFun2} of
the NJL model does not provide any significant advantages as far as
inversion time is concerned over the matrix $M$ of full QCD. The CG
will be as time consuming as in full QCD. The advantage comes because
of another reason. Since the cutoff of the NJL model is at $\ltapprox
1550 \MeV$ and since we will want to look at energies around the pion
mass we are dealing with a ratio of scales $\approx \Lambda / M_\pi
\approx 10$.  Therefore we should expect that lattices of size $16^4$
should be large enough for this purpose. These size lattices can be
simulated in $1 / 4$ of the 64K processor CM-2 supercomputer at SCRI
in a reasonable amount of time. Although this was the original
justification for performing the numerical simulation it turned out
that the numerical results had a more important consequence. As we
will demonstrate in Section 4.1 the numerical results on a given
finite size lattice are in good agreement with the leading order large
N results on the same size lattice, indicating that the $1/N$
corrections are small for the quantities we were able to measure.

The large $N$ results on finite size lattices are obtained by
explicitly performing the four--dimensional finite momentum sums on a
workstation.

Some of the typical parameters of our numerical simulation
are: \hfill\break
\noindent $*$ Trajectory length $\tau = 1$ \hfill\break
\noindent $*$ Step size $d\tau \sim 0.02 - 0.05$ \hfill\break
\noindent $*$ CG residue $10^{-7}$ \hfill\break
\noindent $*$ CG iterations $20-160$ \hfill\break
\noindent $*$ Acceptance rate $> 90 \% $ \hfill\break
\noindent $*$ Mesurements per ``point" $\sim 100$ \hfill\break
\noindent $*$ Autocorelations $2 - 10$ \hfill\break
\noindent $*$ Time for one CG iteration for a $16^4$ lattice on $16$ K
processors of the SCRI CM-2 \hfill\break
\noindent \null\ \ \   $\sim 1.3$ sec

\figure{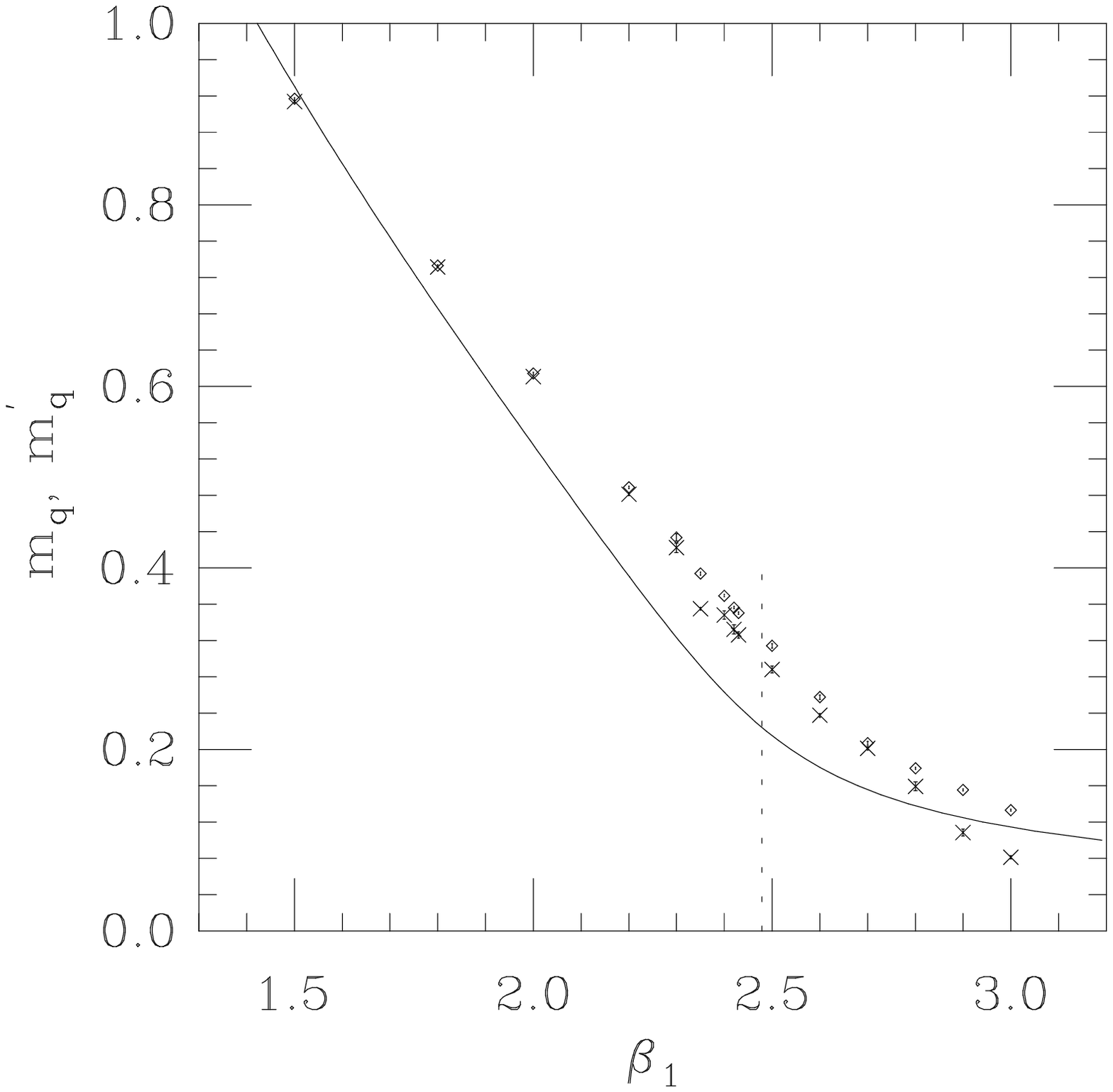}{\three}{\figsizeD}

\figure{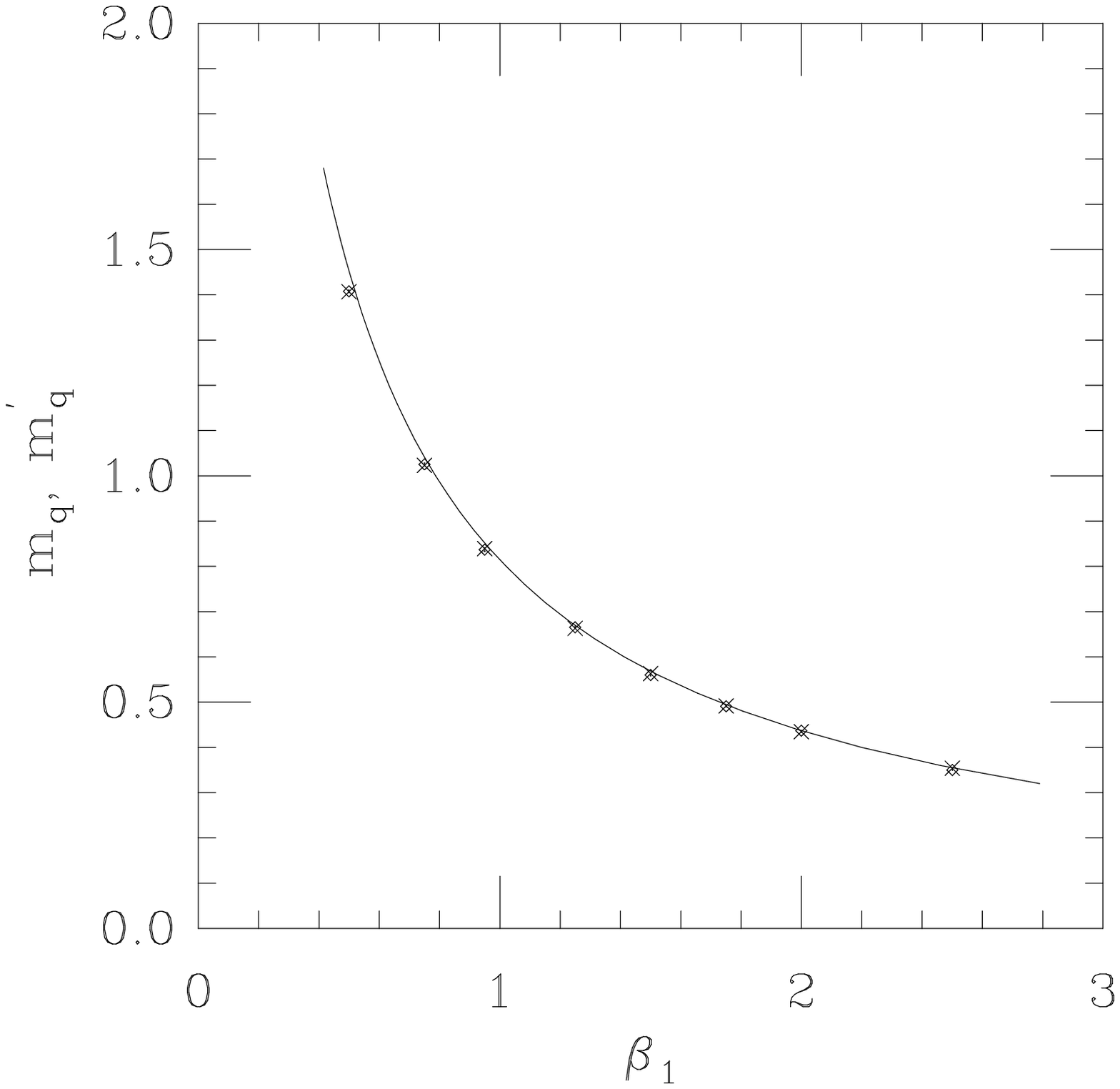}{\four}{\figsizeA}

\subsection{Numerical results and large $N$}

In this section we present our numerical results for $N=2$ and we
compare them with the large $N$ predictions (scaled to $N=2$ when
necessary) on same size lattices.

To check for consistency of the fermionic and auxiliary fields we plot
$m_q$ and $m^{\prime}_q=\{-N <\bar\Psi\Psi> / (2 \beta_1)\} + m_0$ vs.
$\beta_1$ as determined from the numerical simulation.  Because of the
functional identity, eq. \ref{fun_identity}, we should have $m_q =
m^{\prime}_q$. This relation is satisfied nicely.

The agreement with large $N$ of dynamically determined quantities vs.
bare quantities is quite good and helps us to get oriented in the bare
parameter space. This can be seen for various values of the parameters
in Figures 3, 4, 5, and 7a.

The important comparison with large $N$ that will help us get a feel
for the size of the $1/N$ corrections comes from comparisons of
dynamically determined quantities vs. other dynamically determined
quantities. In particular we exchange one of the bare parameters for
$\langle \sigma \rangle$.

\figure{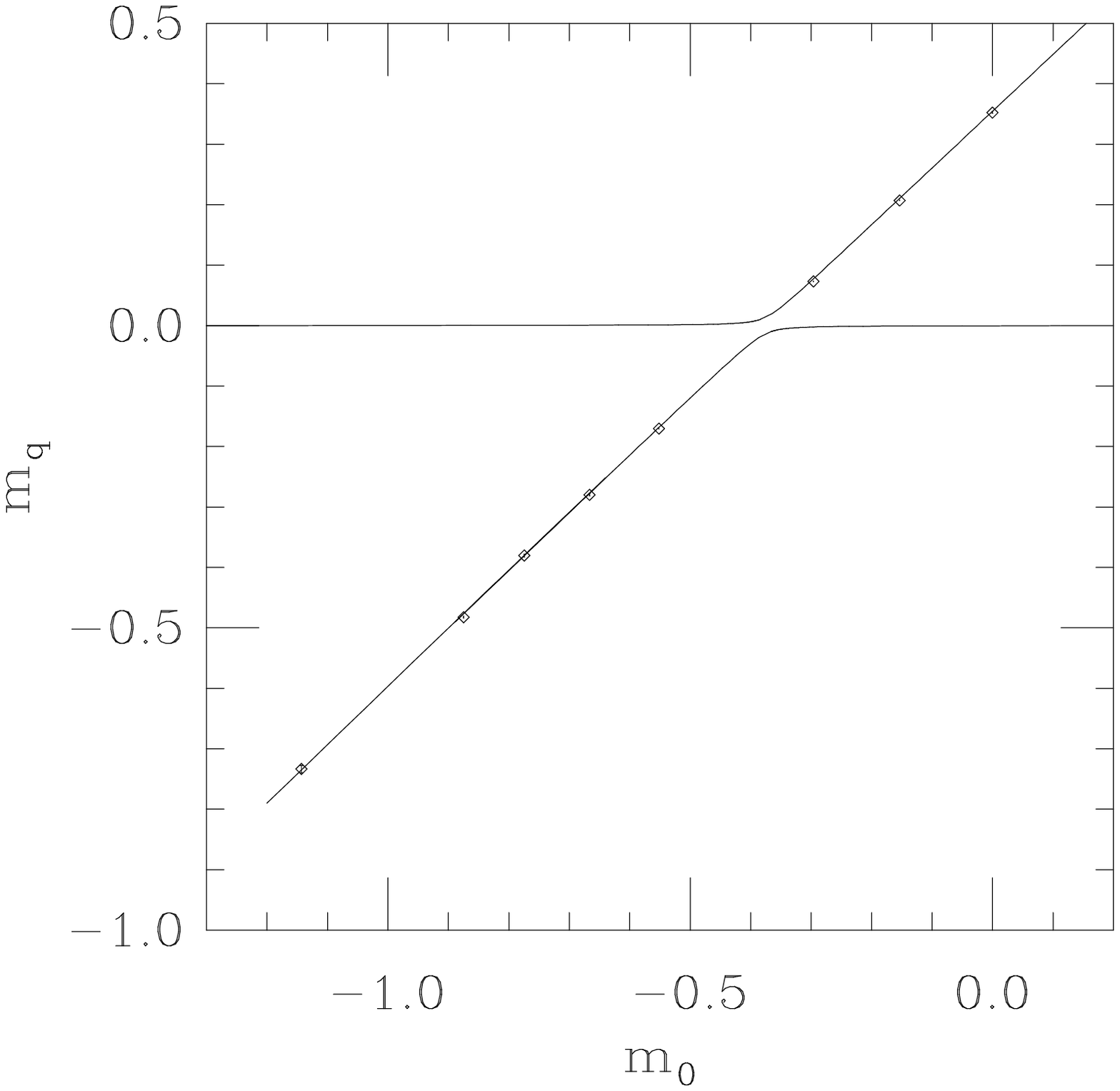}{\five}{\figsizeB}

In Figure 6 we present the $\sigma$ and $\bfm{\pi}$ propagators in
momentum space for ten small momenta and $r=0$. It is from this
figure that we would have to extract $Z_\pi$. As it can be seen the
large $N$ prediction for the same $\langle \sigma \rangle$ as the one
measured in the simulation is in good agreement with the numerical
results. The large $N$ predictions in Figures 6a, 6b, 6c and 6d ``fit''
the numerical results with $\chi^2$ per degree of freedom $0.38$,
$0.67$, $0.32$, $0.42$ respectively. This means that the determination
of $f_\pi= \langle \sigma \rangle \sqrt{N \over Z_\pi}$ as a function
of $\langle \sigma \rangle$ has small $1 / N$ corrections.

\figure{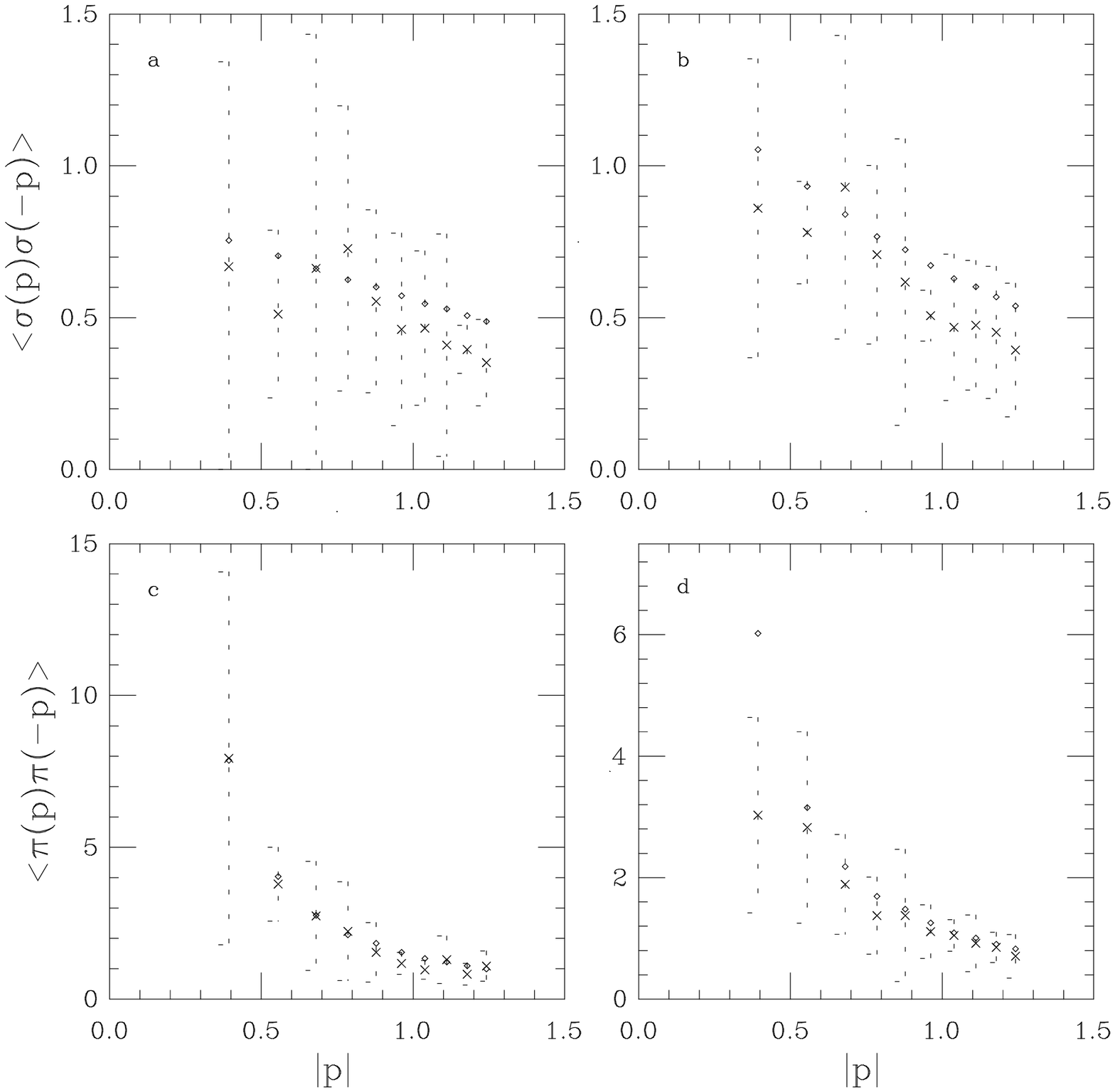}{\six}{\figsizeA}

Another dynamically determined quantity that agrees well with the
large $N$ prediction when plotted vs. $\langle \sigma \rangle$ is the
pion mass $m_\pi$. This plot is shown in Figure 7b. This figure
suggests that the $1/N$ corrections to $m_\pi$ are fairly small.  In
this figure, as well as in figure 7a, $m_\pi$ was not calculated from
the definition \ref{Zpi-mpi_def} but as the imaginary pole of $G_\pi$
(see eq.  \ref{G}).

\figure{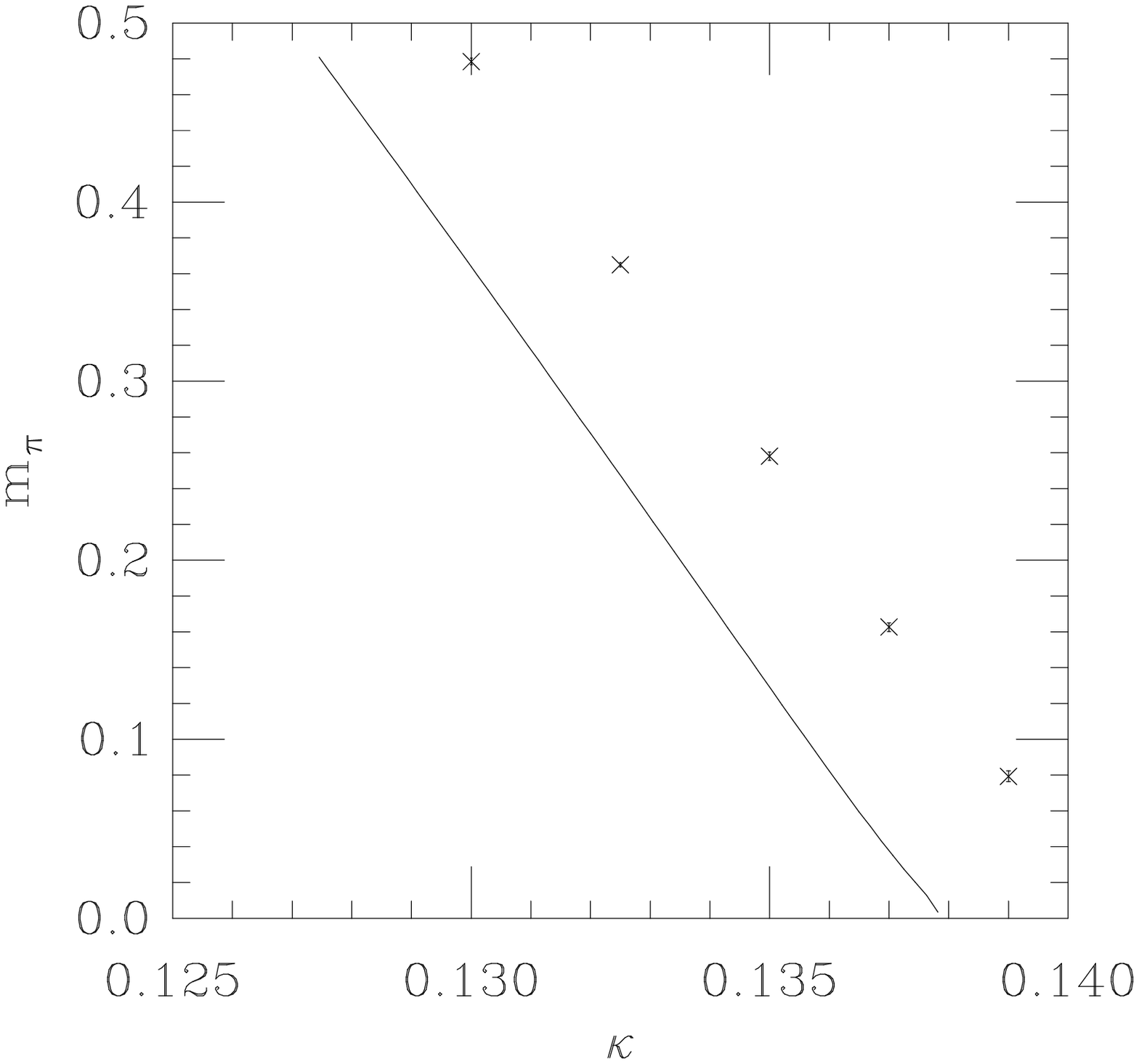}{\sevenA}{\figsizeA}

It should be noted that the good agreement of the large $N$ results
with the numerical simulations is not only present for naive fermions
where the number of species is $16$ times larger and therefore one
would have naively expected the leading order large $N$ expansion to
be a good approximation. It is also present for Wilson fermions ({$r
\ne 0$) as it can be seen from Figures 4, 5, 7a and 7b.

Finally, as discussed in detail in Section 3.2, for $r\neq0$ the
$\sigma$ mass is of order cutoff and therefore very heavy to be able
to measure from the decay of the $\sigma -\sigma$ correlation
function. However, for $r=0$ one would expect to be able to measure
$m_\sigma$. As we will discuss in the next section this is not
possible with the lattice sizes accessible to us. This is unfortunate
since $m_\sigma$ is another very important quantity. However, we
expect the size of the $1/N$ corrections of $m_\sigma$ to be similar
to the ones of $m_\pi$ and therefore fairly small. Also, measurements
of the sigma width were not performed, but, as discussed in Section
3.1, we expect the $1/N$ corrections to the width to be fairly large.

\figure{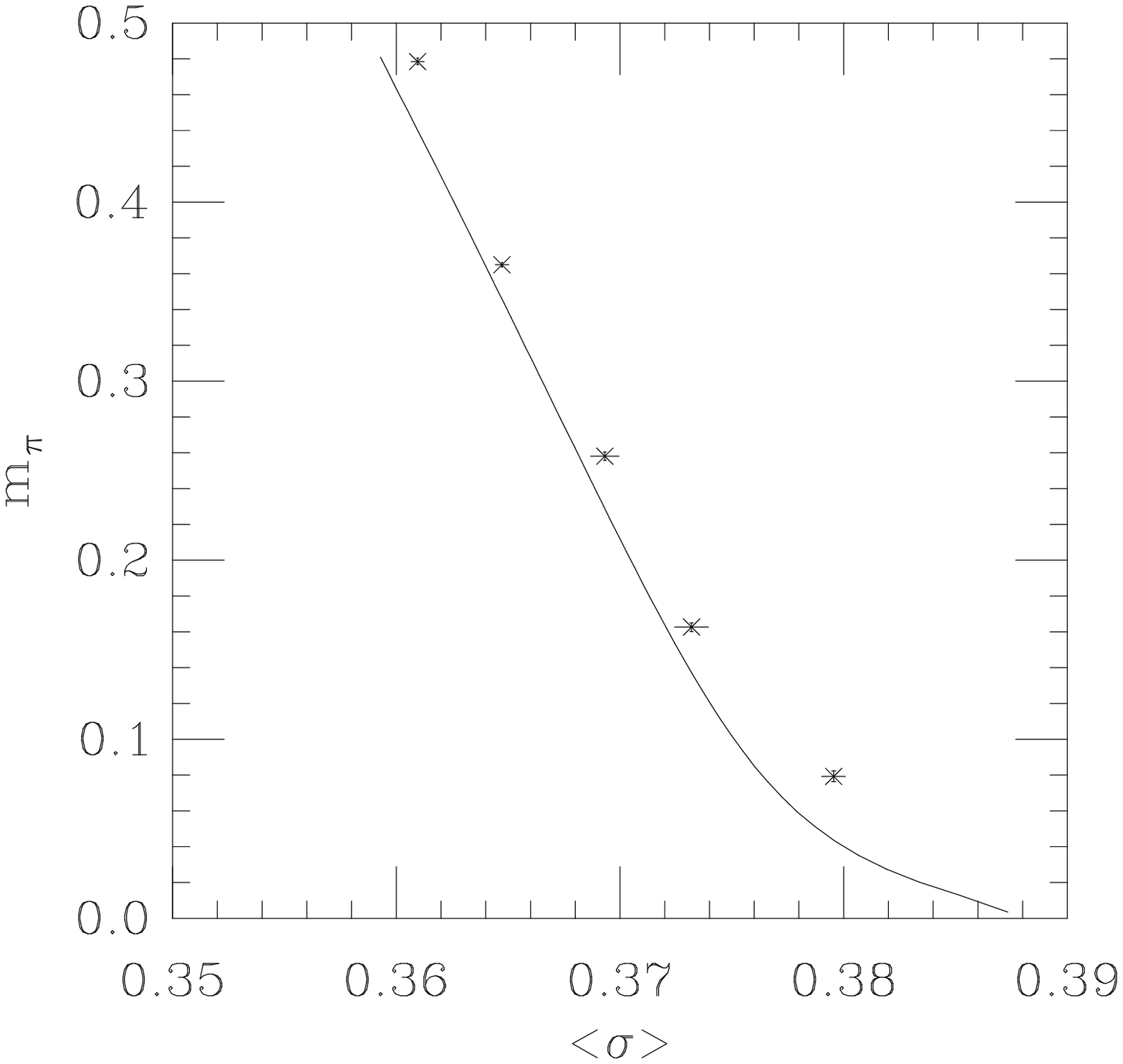}{\sevenB}{\figsizeB}

\subsection{The zero momentum mode of the quarks}

To leading order at large $N$ the matrix $M^\dagger M$ of eq. \ref{PartFun2}
is diagonal in momentum, spin, flavor, and color spaces.
\begin{equation}
M^\dagger M = M_s^\dagger M_s = \sum_\mu \sin^2 p_\mu +
\left[ m_q + r(4 - \sum_\nu \cos p_\mu) \right]^2 + O(1/N)
\label{MdagM}
\end{equation}
The smallest eigenvalue of this matrix is $m_q^2$ and corresponds to
the $p=0$ matrix element. This in turn corresponds to the zero
momentum modes of the quarks which from now on we will simply refer to
as ``zero modes''.

\figure{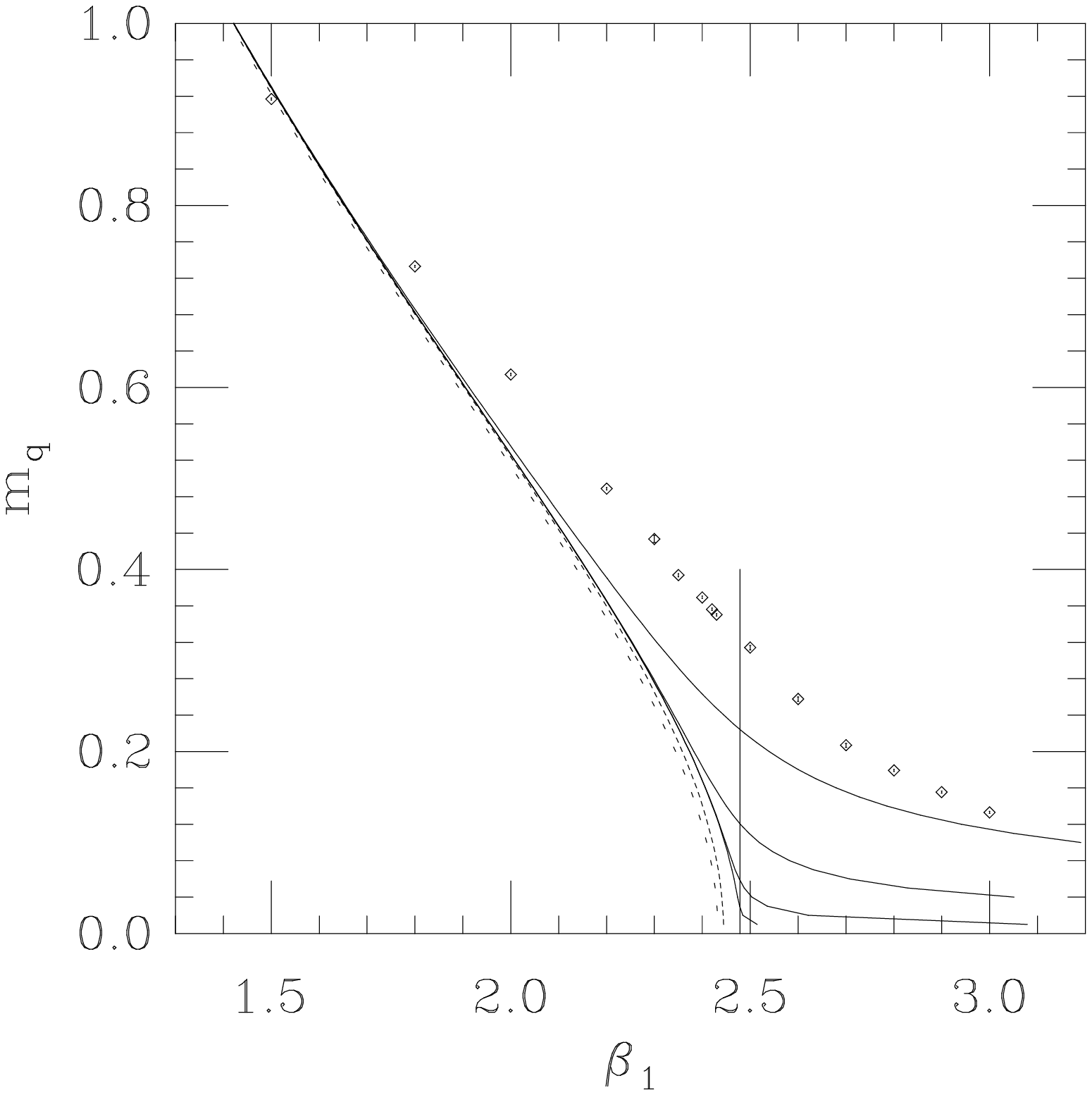}{\eight}{\figsizeA}

For small $m_q$ the condition number $c$ of this matrix is:
\begin{eqnarray}
c &\approx& {4 \over m_q^2} \ \ {\rm for} \ \ r=0 \nonumber \\
c &\approx& {64 r^2 \over m_q^2} \ \ {\rm for} \ \ r=1
\label{cond_num}
\end{eqnarray}
A large condition number will make the inversion of $M^\dagger M$ very
slow. An important observation can be made by noticing the dependence
of the condition number on $r$.  This suggests that performing the
simulation with smaller $r$ will yield a quite faster inversion. It is
possible that this may also be the case for QCD.

\figure{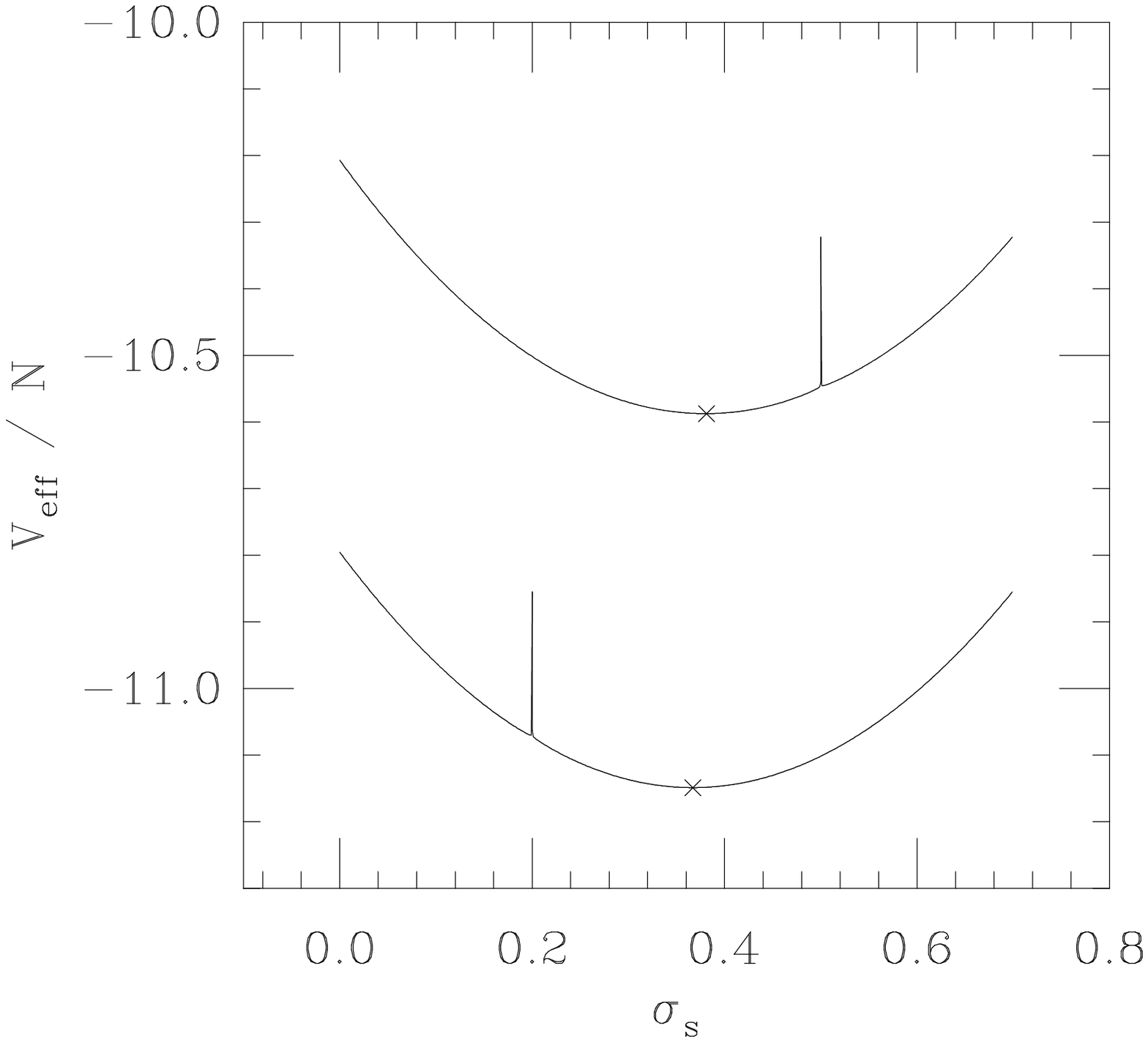}{\nine}{\figsizeA}

But the unwelcomed effect of the zero modes on a finite lattice is not
limited to large inversion times. Because on a finite lattice their
effects are not suppressed by the measure but instead by an inverse
volume factor, it turns out that in certain cases they severely
obscure the physics.

\figure{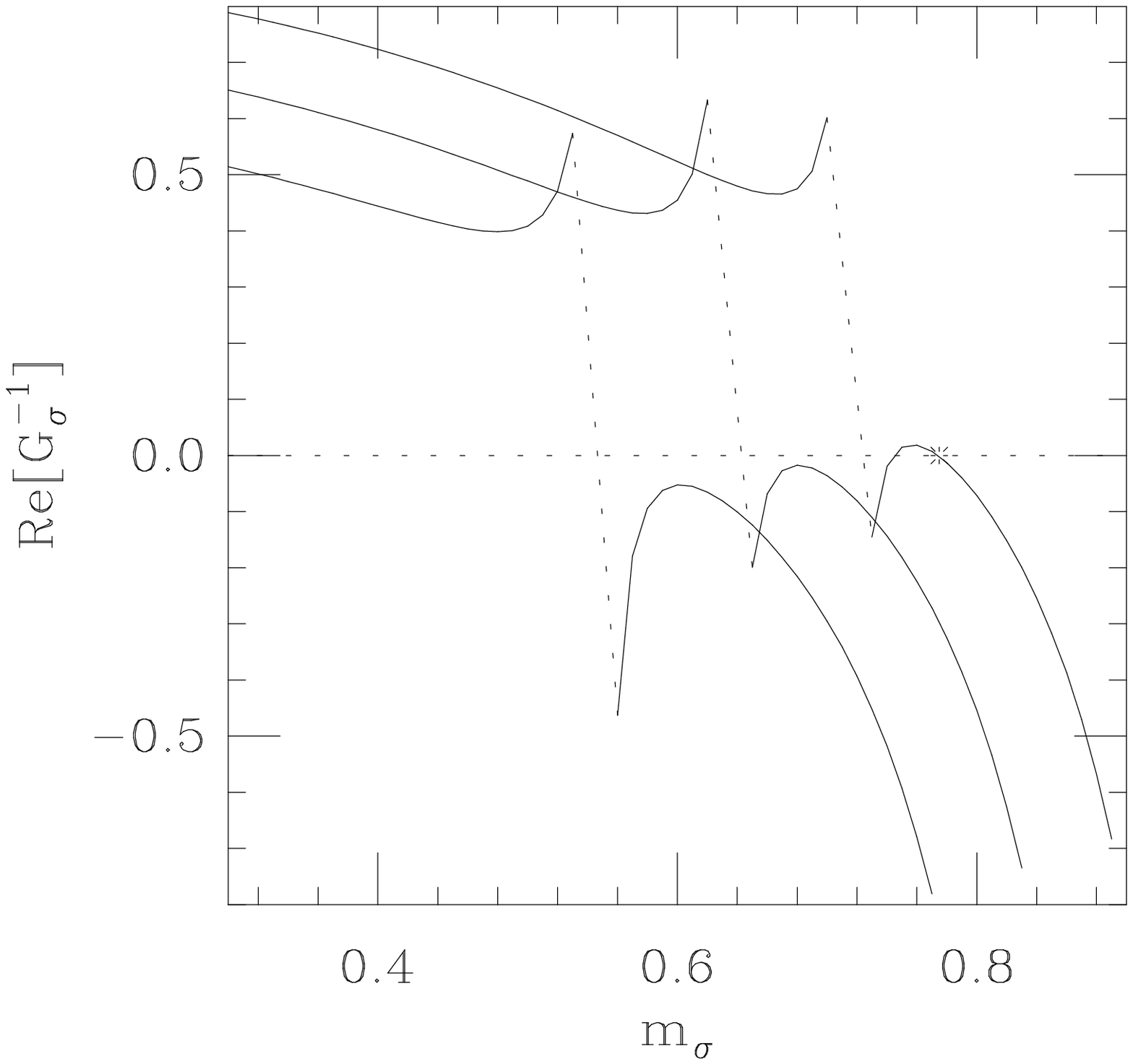}{\tenA}{\figsizeA}

In Figure 8 we plot $m_q = \langle \sigma \rangle $ vs. $\beta_1$ for
$r=0$, $m_0=0$, $L_x=8$, and $L_t=16$ (same as Figure 3). By simply
looking at the numerical results we would not only be unable to estimate
the critical point but also we would be unable to see any indication
of a phase transition. The large $N$ result on the same size lattice
also has the same problems. As we increase the lattice size in the
large $N$ calculation (solid lines from top to bottom) we see that a
picture of an order parameter slowly materializes. At $ L_x=64 $, $
L_t=64 $ a fairly good prediction of the large $N$ infinite volume
critical point is achieved. If we now do the same large $N$
calculation but neglect from the momentum sums the zero modes, we
obtain as a result the two dotted lines for $L_x=8$, $L_t=16$ and
$L_x=16$, $L_t=16$ (from left to right). We see that neglecting the
zero modes on an $L_x=8$, $L_t=16$ lattice gives very similar results
as the ones obtained on a $L_x=64$, $L_t=64$ lattice with the zero
modes included.

\figure{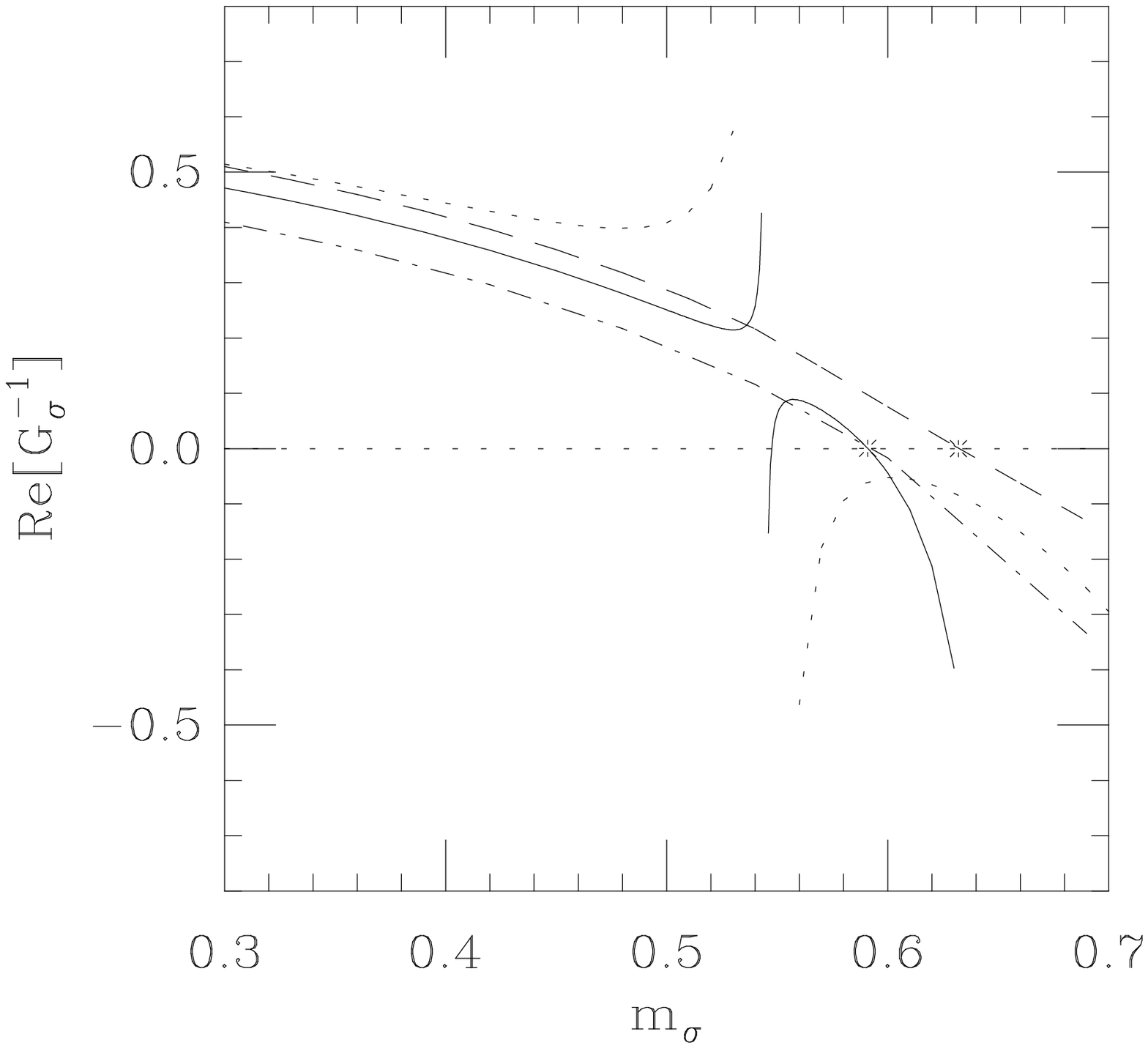}{\tenB}{\figsizeA}

If we plot the effective potential of eq. \ref{eff_pot} vs. $\sigma_s$
for a finite lattice we will obtain a result as in Figure 9.  The
``spike'' is a result of the presence of the zero mode on the lattice
sum and extends to infinity. The presence of this spike may create
thermalization problems if the initial configuration is chosen on the
``wrong'' side of the ``spike''. Of course the ``width'' of the spike
is negligible and therefore this problem may not be important.

As mentioned at the end of the previous section, although one would
expect to be able to measure the $\sigma$ mass in the $r=0$ case we
were not able to do so. Large $N$ provides an explanation of this
unexpected problem. In Figure 10a we plot the real part of the inverse
$\sigma$ propagator, eq. \ref{G}, for a finite lattice with $L_x =16$,
$L_t=16$ and external momentum set to $q=\{i m_\sigma,0,0,0\}$. The
sigma mass should be obtained at the zero of this function. We see
that because of the presence of a discontinuity we do not obtain a
root until $m_\sigma$ becomes heavy. The presence of this
discontinuity is again due to the zero modes.

In Figure 10b we plot the left most case of Figure 10a (dotted line).
As we already mentioned there is no zero. If we increase the lattice
size to $L_x =32$, $L_t=32$ (solid line) we see that a zero develops.
The infinite volume result from the asymptotic expansion (dashed line)
has a zero nearby and no discontinuity since the zero modes are fully
suppressed.  If on the $L_x =16$, $L_t=16$ lattice we now exclude the
zero modes (dot--dash line) we see that the discontinuity disappears
and a zero very close to the infinite volume result is obtained.

\subsection{The $m_\pi=0$ line on a finite lattice at large $N$}

The $m_\pi=0$ zero line is of particular importance since it is there
that the continuum chiral limit is obtained.  As discussed in Section
3.2 this limit is obtained in the part of the $m_\pi=0$ line that
corresponds to small quark mass. That region was presented in Figures
2a, 2b for $-0.2 \leq m_q \leq 0.3$. If, for the same values of $m_q$,
the zero pion mass line is calculated on a finite volume one finds
that the corresponding ranges of $\kappa$ and $\beta_1$ change by
almost an order of magnitude. To be more specific, for $r=1$ the point
of the the zero pion mass line corresponding, for example, to
$m_q=0.015$ has $m_0= -2.75$ and $\beta_1 = 0.34$ at infinite volume
(see Figure 2a).  The same point on an $L_x = 8$, $L_t = 16$ lattice
has $m_0=-0.37$ and $\beta_1=2.51$. This change is fairly unusual and
it may appear as if there is a contradiction between the finite and
infinite volume results. In particular, in the infinite volume work of
\cite{Aoki} no point with $m_\pi=0$ was found for $\beta_1 > 1.41$. To
clarify this issue we plot in Figure 11a the $m_\pi=0$ line for an
$L_x = 8$, $L_t = 16$ lattice for $-\infty < m_q < +\infty$. This line
is calculated as follows: For a given value of $m_q$ we use the finite
volume version of eq.  \ref{mpi0_line}
\begin{equation}
m_0 {1 \over L_x^3 L_t} \sum_k {1 \over g(k,m_q)} + r {1 \over L_x^3
L_t} \sum_k {w(k) \over g(k,m_q)} =0
\label{mpi0_line_fnt}
\end{equation}
to calculate $m_0$. Next, using this $m_q$ and $m_0$ we calculate
$\beta_1$ by combining equations \ref{quark_mass} and \ref{sp_eq}. The
finite volume version of this is:
\begin{equation}
\beta_1 = {2 N \over m_q - m_0} {1 \over L_x^3 L_t} \sum_p {m_q + r
w(p) \over g(p,m_q)} \fs
\label{sp_eq_fnt_vol}
\end{equation}

At point ``1" of Figure 11a $m_q$ is very large and positive. As $m_q$
decreases, crosses zero and tends to very large negative values, we
transverse the whole solid line from ``1" to ``2" to ``3" $\cdots$ to
``12" where $m_q$ is very large and negative.  The ``prongs" of this
figure, points ``3", ``5", ``7", ``9", and ``11", extend all the way to
$\beta_1=\infty$ where they correspond to $m_q = 0, -2, -4, -6, and -8$,
respectively.  This singular behavior originates from the terms of the
momentum sum in eq.  \ref{sp_eq_fnt_vol} that correspond to the
origins of the ``Brillouin zones" of the 16 species (see Table 1).
When these terms are neglected from the calculation (the contribution
of these terms disappears in the infinite volume limit) we obtain the
dotted line of Figure 11a.

\figure{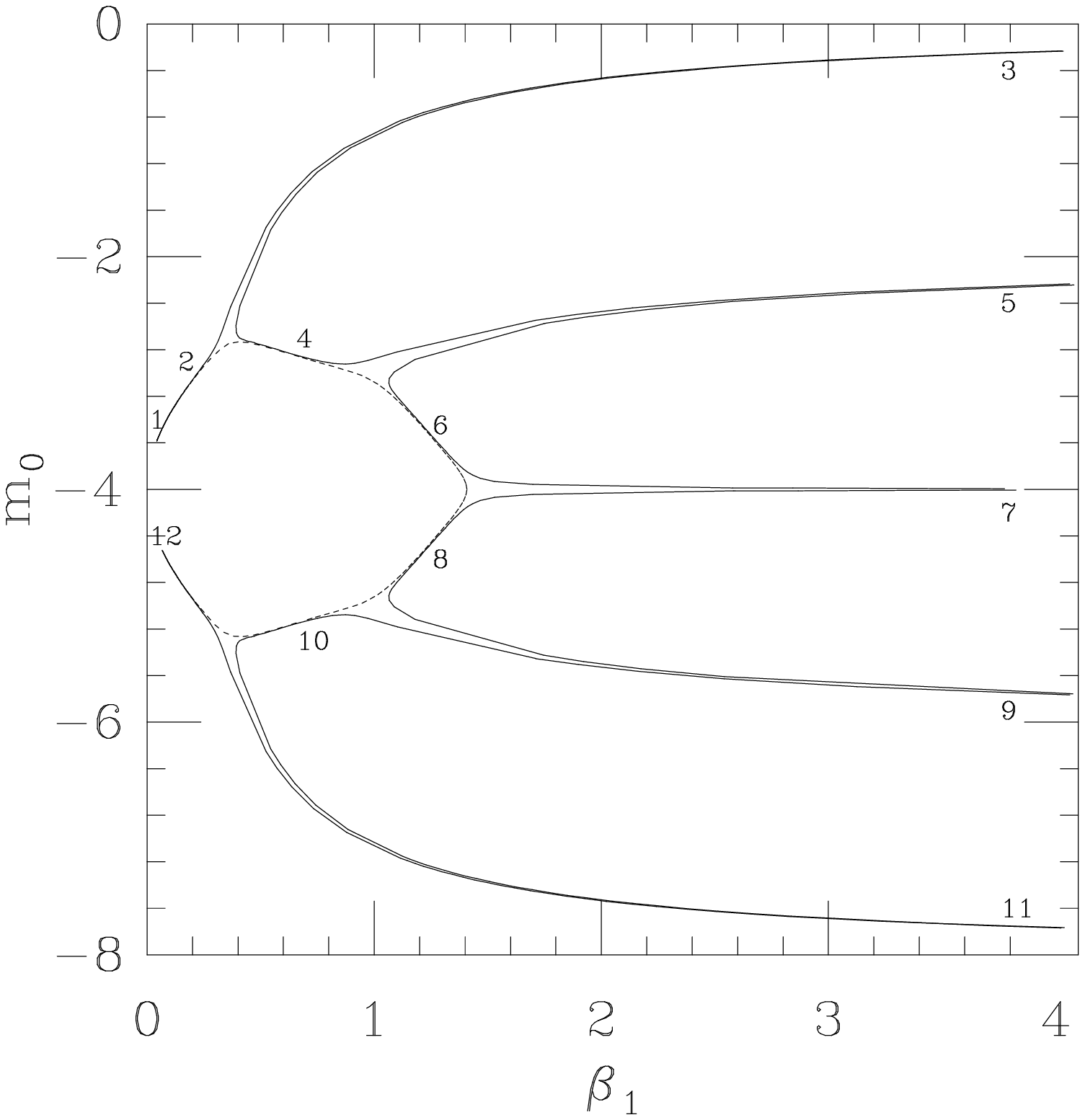}{\elevenA}{\figsizeA}

To see more clearly how this behavior
develops as we increase the volume we concentrate, as an example, on
the ``prong" that corresponds to $m_q=0$ since it is this ``prong"
that is most interesting in the recovery of continuum physics.  A
blown up picture of this ``prong" is plotted in Figure 11b. The outer
solid lines correspond to the $L_x = 8$, $L_t = 16$ lattice. Moving
inwards the solid lines correspond to an $L_x = 16$, $L_t = 16$
lattice and an $L_x = 32$, $L_t = 32$ lattice. The dashed line
corresponds to the $L_x = 8$, $L_t = 16$ lattice but with the singular
terms removed.  The dotted line is the infinite volume result of
Figure 2a for $-0.1 \leq m_q \leq 0.1 $. Because the singularities are
only quadratic they have disappeared in the infinite volume limit. The
circle corresponds to the infinite volume $m_q=0, m_\pi=0$ point.  As
the volume increases the whole ``prong" is slowly mapped down to the
infinite volume result of the dotted line. It is also very interesting
to notice how close the infinite volume result is to the $L_x = 8$,
$L_t = 16$ lattice result with the singular terms removed.  Similar
behavior is expected for the other ``prongs" as well.

\figure{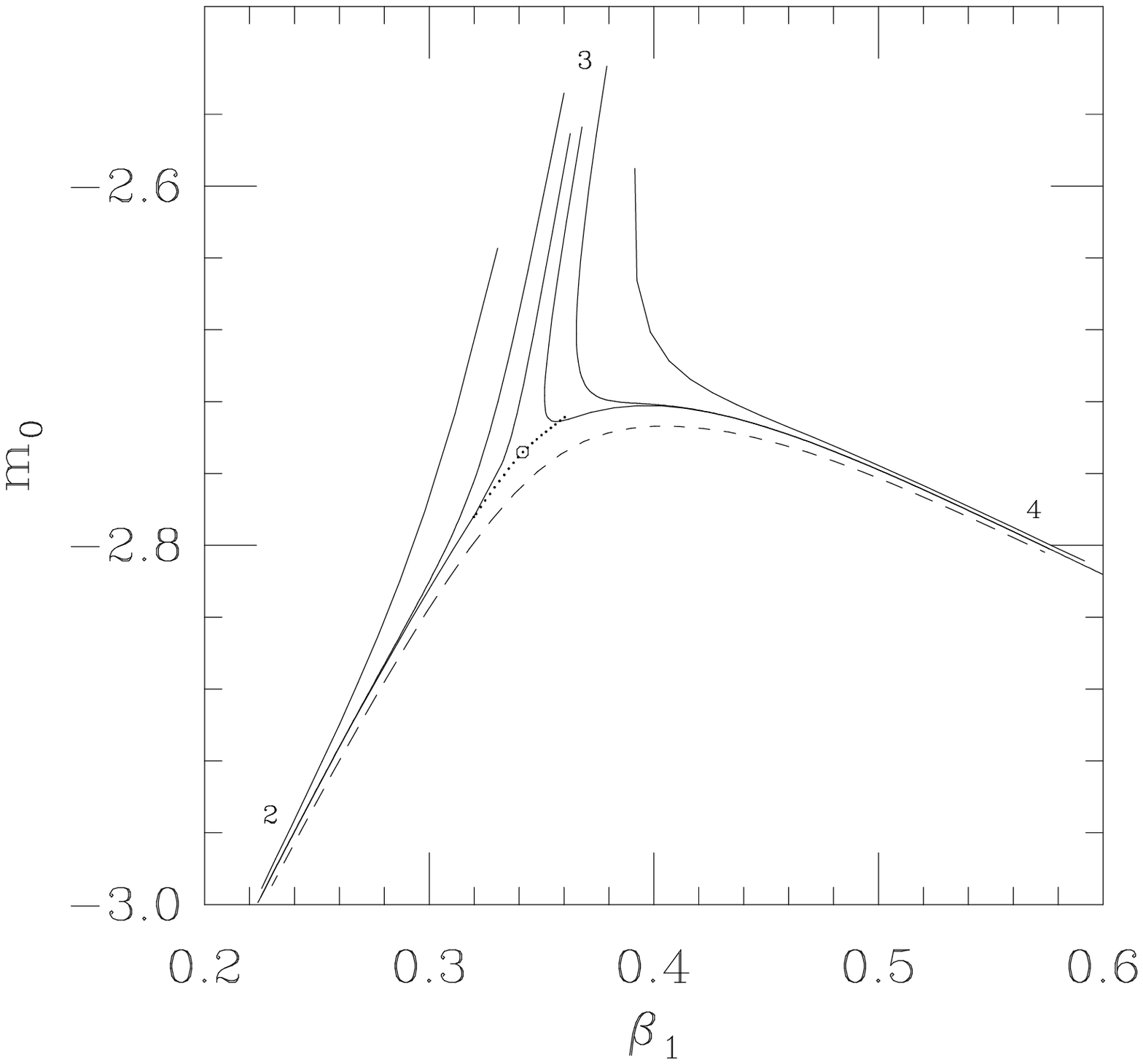}{\elevenB}{\figsizeA}

In \cite{Aoki} it was shown that the $m_\pi=0$ line is the phase line
that separates the parity-flavor symmetric phase ($\langle|{\bfm
\pi}|\rangle=0$) with the spontaneously broken parity-flavor symmetry
phase ($\langle|{\bfm \pi}|\rangle\neq 0$). We have confirmed that
this is also the case on the finite lattice. The region inside the
octopus--like graph of Figure 11a has ($\langle|{\bfm \pi|}\rangle\neq
0$).

Finally, notice that the numerical results for $m_\pi$ of Figure 7a
indicate that on a finite lattice $m_\pi$ can be made quite small.
The place where this happens is predicted by large $N$ quite well and
it corresponds to a point on ``prong" ``3" that is quite far from
where the corresponding point of the infinite volume would be
(somewhere on the dotted line on the ``inside" of the ``prong").

\section{Conclusion and Summary}

In this paper we have found that the lattice version of the NJL model
is an excellent toy model to investigate issues related to lattice
QCD. We have used the large $N$ approximation to leading order in
$1/N$ to obtain non perturbative analytical results over almost the
whole parameter range. By using numerical simulations of the model we
estimated that for most of the quantities we considered the $1/N$
corrections are small.

The main results of our investigation are listed below:

\medskip
\noindent
{\bf 1)} With Wilson fermions we obtain at large $N$ analytical
expressions of the pion mass ($m_\pi$) and quark mass ($m_q$) in
lattice units as functions of the bare parameters of the model. We are
then able to make exact statements regarding the approach to the
continuum chiral limit. The ``phase diagram" is presented in Figure 2.
This may provide an insight on how the retrieval of the continuum
chiral limit is achieved in QCD.

\medskip
\noindent
{\bf 2)}
At large $N$ and for Wilson fermions the $\sigma$ particle has mass
proportional to the cutoff. Our analysis traces this fact to two
related reasons. First, although the Wilson term has raised the masses
of the doublers to the cutoff, it has not decoupled them from the
theory.  Through vacuum polarization these contribute to the $\sigma$
self energy and raise its mass. Second, although the Wilson term has
not altered the low frequency behavior of the propagating quark, it
has however altered its high frequency behavior. Again through vacuum
polarization the high frequency modes contribute to the $\sigma$ self
energy and also raise its mass. Such a phenomenon may also be
responsible for the difficulty in observing a $\sigma$ particle in
numerical simulations of QCD with dynamical Wilson fermions.

\medskip
\noindent
{\bf 3)} The numerical simulation is performed on finite lattices. For
naive fermions one would expect to be able to see some indication of
the chiral phase transition as well as a $\sigma$ particle. However,
by simply looking at the graph of the vacuum expectation value vs. the
coupling (see fig. 8) on an $8^3 \times 16$ lattice one can not see any
indication of a phase transition. Also the $\sigma$ particle in a
$16^4$ lattice is either non existent or too heavy to be measured.
Both of these unexpected results can be explained at large $N$. The
reason is traced to the existence of zero quark momentum modes that on
a finite lattice are not sufficiently suppressed. The zero modes
besides obscuring some of the physics are also probably partially
responsible for the large inversion times in the HMC algorithm. To
leading order at large $N$ the smallest eigenvalue of the matrix that
has to be inverted is $m_q^2$ and corresponds to the zero quark
momentum mode. For small $m_q$, the condition number of the matrix is
$4/m_q^2$ for $r=0$ and $64 r^2 / m_q^2$ for $r=1$.  An important
observation can be made by noticing the dependence of the condition
number on $r$.  This suggests that performing the simulation with
smaller $r$ will yield a quite faster inversion. It is possible that
this may also be the case for QCD.

\medskip
\noindent
{\bf 4)} The observables measured in the numerical simulation (chiral
condensate, vacuum expectation value, pion wave function
renormalization constant, pion mass) have values that are in good
agreement with leading order large $N$.  This provides a quantitative
prediction for the size of the $1/N$ corrections. In agreement with
the large $N$ predictions discussed in 2 and 3 above, the $\sigma$
mass was very heavy to give a good signal and was not measured. Also,
measurements of the sigma width were not performed, but, as it will be
discussed in 5 below, we expect the $1/N$ corrections to the width to
be large.

\medskip
\noindent
{\bf 5)} For naive fermions we calculate at large $N$ and with
$M_\pi=140 \MeV$ the $\sigma$ mass ($M_\sigma$), the $\sigma$ width
($\Gamma_\sigma$) and the quark mass ($M_q$)in physical units as
functions of the cutoff. By setting $M_q = 310 \MeV$ we find $M_\sigma
= 726 \MeV$, $\Gamma_\sigma=135 \MeV$, and $\Lambda = \pi / a = 1150
\MeV$. $M_\sigma$ is consistent with phenomenological expectations and
$\Lambda$ is consistent with the expectation that the cutoff should be
close and below the mass of the lightest glueball ($1550 \MeV$). The
width however is underestimated. The reason is traced to the fact that
to leading order in large $N$ the width receives contributions only
from the quark bubble and not from the pion bubble because the pion
bubble is of order $1/N$. Because the phase space available for the
$\sigma$ to decay to two quarks is much smaller than the phase space
to decay to two pions the pion loop contribution although of order
$1 / N$ is probably more important than the quark loop contribution.

\medskip
\noindent
{\bf 6)} The above result can also be used to make an interesting
observation.  If the Higgs sector is the low energy effective field
theory of a NJL model with exactly the same parameters as the low
energy QCD except for $M_\pi=0$ and $F_\pi=246 \GeV$, then we find
$M_\sigma = 1915\GeV$.  This corresponds to $m_\sigma = 2$ where one
would expect very large deviations from the low energy behavior of
scattering cross sections.  Although we have not calculated these
deviations the value of the width serves as an indication of their
size. In a way, departure from low energy behavior will be signaled by
an increasing width of the $\sigma$ to two quark decay. At $m_\sigma
\approx 2$ the width is already fairly large.

There are some interesting issues relevant to lattice work that have
not been considered in this paper.  It would be important to calculate
the three and four point vertices and therefore be able to calculate
scattering amplitudes and their departure from low energy behavior as
well as the $1/N$ corrections to the width. It would also be
interesting to study the NJL model at finite temperature and
investigate the finite temperature transition in connection with the
approach to the continuum chiral limit. Finally it would be important
to include vector meson couplings (see, for example,
\cite{Ebert-Volkov}, \cite{Ebert-Reinhard}) and confirm that for the
case of Wilson fermions the vector meson masses scale appropriately
and do not become of the order cutoff as the $\sigma$ particle does.

\section*{Acknowledgments}

We would like to thank U.M. Heller and R. Edwards for useful
discussions concerning the subject of this paper. This research was
supported in part by the DOE under grant $\#$ DE-FG05-85ER250000 and
$\#$ DE-FG05-92ER40742.

\hfill

\if \PRD N
\eject
\fi

\section*{Appendix A}

Below we give the lattice constants, defined in eq. \ref{anm} and
\ref{const_r}, that appeared in our analysis of Wilson fermions.

\noindent{\bfm{r=1}}
\begin{eqnarray}
a_{0,1} &=& 0.0854    \nonumber \\
a_{1,1} &=& 0.2347    \nonumber \\
a_{1,2} &=& 0.0260    \nonumber \\
a_{2,2} &=& 0.0597    \nonumber \\
a_{2,3} &=& 0.0086    \nonumber \\
a_{3,3} &=& 0.0165    \nonumber \\
r_1     &=&-0.0119    \nonumber \\
s_1     &=& {1 \over 16 \pi^2}
\label{anm_r1_numbers}
\end{eqnarray}
Although they were never used, we have also calculated the constant
$R_3$ defined in eq. \ref{lat_consts} and also few more of the
$a_{n,m}$'s.
\begin{eqnarray}
R_3     &=&-0.0016    \nonumber \\
a_{4,4} &=& 0.0050    \nonumber \\
a_{4,3} &=& 0.0470    \nonumber \\
a_{6,4} &=& 0.0388
\label{anm_r1_var}
\end{eqnarray}
\noindent{\bfm{r=0.1}}
\begin{eqnarray}
a_{0,1} &=& 0.5373    \nonumber \\
a_{1,1} &=& 2.0721    \nonumber \\
a_{1,2} &=& 1.3881    \nonumber \\
a_{2,2} &=& 5.8935    \nonumber \\
a_{2,3} &=& 6.1127    \nonumber \\
a_{3,3} &=& 25.633    \nonumber \\
r_1     &=&-0.3818    \nonumber \\
s_1     &=& {1 \over 16 \pi^2}
\label{anm_r1}
\end{eqnarray}

The asymptotic expansions of the integrals $J_{0,1}(m_q)$ and
$J_{1,1}(m_q)$ defined in eq. \ref{J-I0_r} to order $m_q^2$ are:
\begin{eqnarray}
J_{0,1}(m_q) &=& a_{0,1} - 2 m_q r a_{1,2} + m_q^2 [ 4 r^2 a_{2,3} +
r_1 + s_1 + s_1 \log m_q^2] + O(m_q^3) \nonumber \\
J_{1,1}(m_q) &=& a_{1,1} - 2 m_q r a_{2,2} + m_q^2 [4 r^2 a_{3,3} -
a_{1,2}] + O(m_q^3)
\label{AJ1}
\end{eqnarray}
The leading order term of the asymptotic expansion of $J_{0,2}(m_q)$
defined in eq. \ref{J-I0_r} is:
\begin{equation}
J_{0,2}(m_q) = - r_1 - s_1 - s_1 \log m_q^2 + O(m_q)
\label{AJ2}
\end{equation}
%


\if \epsfpreprint N

\eject

\section* {Figure Captions.}

\noindent{\bf Figure 1:} \one

\noindent{\bf Figure 2a:} \twoA

\noindent{\bf Figure 2b:} \twoB

\noindent{\bf Figure 3:} \three

\noindent{\bf Figure 4:} \four

\noindent{\bf Figure 5:} \five

\noindent{\bf Figure 6:} \six

\noindent{\bf Figure 7a:} \sevenA

\noindent{\bf Figure 7b:} \sevenB

\noindent{\bf Figure 8:} \eight

\noindent{\bf Figure 9:} \nine

\noindent{\bf Figure 10a:} \tenA

\noindent{\bf Figure 10b:} \tenB

\noindent{\bf Figure 11a:} \elevenA

\noindent{\bf Figure 11b:} \elevenB

\fi

\bigskip

\if \PRD Y

\section* {Table Captions.}

\noindent{\bf Table 1:} The masses of the $16$ species for small $r$.

\fi

\end{document}